\documentclass[prl,aps,twocolumn,showpacs,showkeys,reprint,superscriptaddress,amsmath,amssymb]{revtex4-1}
\usepackage{graphicx}
\usepackage{hyperref}
\hypersetup{
bookmarks=false,
anchorcolor=blue,
colorlinks=true,
citecolor=blue,
urlcolor=blue,
linkcolor=blue}
\usepackage{xcolor}
\usepackage{bbold}
\usepackage[noindentafter]{titlesec}
\titleformat{\paragraph}[runin]{\normalfont\normalsize\itshape}{\theparagraph}{0pt}{}[]

\newcommand{\imi}{\mathrm{i}}

\begin{document}

\title{Steady-state generation of Wigner-negative states in 1D resonance fluorescence}

\author{Fernando Quijandr\'{\i}a}
\affiliation{Microtechnology and Nanoscience, MC2, Chalmers University of Technology, SE-412 96
G\"oteborg, Sweden}
\author{Ingrid Strandberg}
\affiliation{Microtechnology and Nanoscience, MC2, Chalmers University of Technology, SE-412 96
G\"oteborg, Sweden}
\author{G\"oran Johansson}
\affiliation{Microtechnology and Nanoscience, MC2, Chalmers University of Technology, SE-412 96
G\"oteborg, Sweden}

\date{\today}

\begin{abstract}
In this work we demonstrate
numerically that the nonlinearity provided
by a continuously driven
two-level system (TLS) allows for the generation
of Wigner-negative states of the electromagnetic field
confined in one spatial dimension.
Wigner-negative states, a.k.a.\ Wigner nonclassical states, are desirable for quantum information protocols
beyond the scope of classical computers.
Focusing on the steady-state emission from the TLS,
we find the largest negativity at the drive strength where the
coherent reflection vanishes.
\end{abstract}

\maketitle

\paragraph{Introduction.---}\hspace{0pt}
\noindent In this Letter we present the calculation of the Wigner
function of one-dimensional
($1$D) steady-state resonance fluorescence.
Resonance fluorescence, the spontaneous emission from a two-level system (TLS)
driven by a resonant electromagnetic field~\cite{Walls}, is
one of the simplest theoretical models for studying the 
light-matter interaction.
Despite its basic structure, this model
exhibits very rich phenomena, including photon
antibunching~\cite{Kimble1977}, squeezing in the scattered field of a very weak
drive~\cite{Walls1981}, and an inelastic scattering spectrum (Mollow
triplet) for a strong drive~\cite{Mollow1969}.

The Wigner function~\cite{Wigner1932, Hillery1984, Gardiner},
a quasi-probability distribution which allows for a
description of quantum mechanics in phase space, recently
gathered
relevance in
the context of continuous variable (CV) quantum
information~\cite{Braunstein1999, Gu2009}.
This is because it allows to discern the class of states
necessary for achieving a quantum advantage
over classical simulations.
This class corresponds to Wigner-negative states,
i.e, states characterized by a \textit{negative}
Wigner function~\cite{Mari2012, Veitch2013, Rahimi2016}.
Assuming a coherent drive,
a nonlinearity
is required in order to generate Wigner
negativity.

The TLS is a nonlinear medium interacting with the incoming
radiation, causing the reflection 
(transmission)
to be dependent
on the intensity of the radiation.
In order  to avoid a spatial mode mismatch between
incoming and scattered fields, it is desirable to confine the emission
to a single spatial dimension. Artificial 1D systems provide an ideal testbed
for studying the light-matter interaction
due to its enhancement
as a consequence of the confinement~
\cite{Chang2007,Muller2007,Astafiev2010, Peropadre2013, Hoi2011,Schulte2015, Gu2017}.

Here we study the resonant scattering
from a TLS in front of a mirror.
The mirror serves two purposes. 
First, it avoids the loss of information due to an unobserved channel: a transmitted
component. Second, and 
more important, it has been
demonstrated that in this configuration it is possible to
cancel the coherent component of the scattered field~\cite{Koshino2012,Hoi2015}.
It is known that 
this setup allows for
single-photon generation when the TLS is driven with a pulse.
Prominent examples of this are superconducting circuits
\cite{Eichler2011, Yin2013, Sankar2016, Forn-Diaz2017} and quantum dot
\cite{Matthiesen2012,He2013, Lodahl2015, Ding2016} setups.
Single photons have no classical counterpart and are correspondingly
characterized by a negative Wigner function.
Still, the question of steady-state emission
from a \textit{continuously driven} TLS remains.
In contrast to the
pulsed scheme for which the exponential decay of the TLS
gives a known probability of detecting the emitted photon in time,
we lack this information when driving continuously.
The uncertainty in the emission time of the photon
enhances the role of the vacuum in the emitted state.
Additionally, Fock states other than the vacuum and a single-photon
may contribute to the output field.
In this Letter we explore if the nonlinearity
of a continuously driven TLS
 suffices to generate Wigner nonclassical states of 
light.
Furthermore, in contrast
to nondeterministic nonlinear operations
commonplace in quantum optics such as
photon substraction~\cite{Kim2008} or
cubic phase gate
approximations~\cite{Yukawa2013,Marshall2015,Arzani2017}, we explore steady-state
and deterministic generation of this class of states.

\begin{figure}[t]
\begin{center}
\includegraphics[width=0.85\columnwidth]{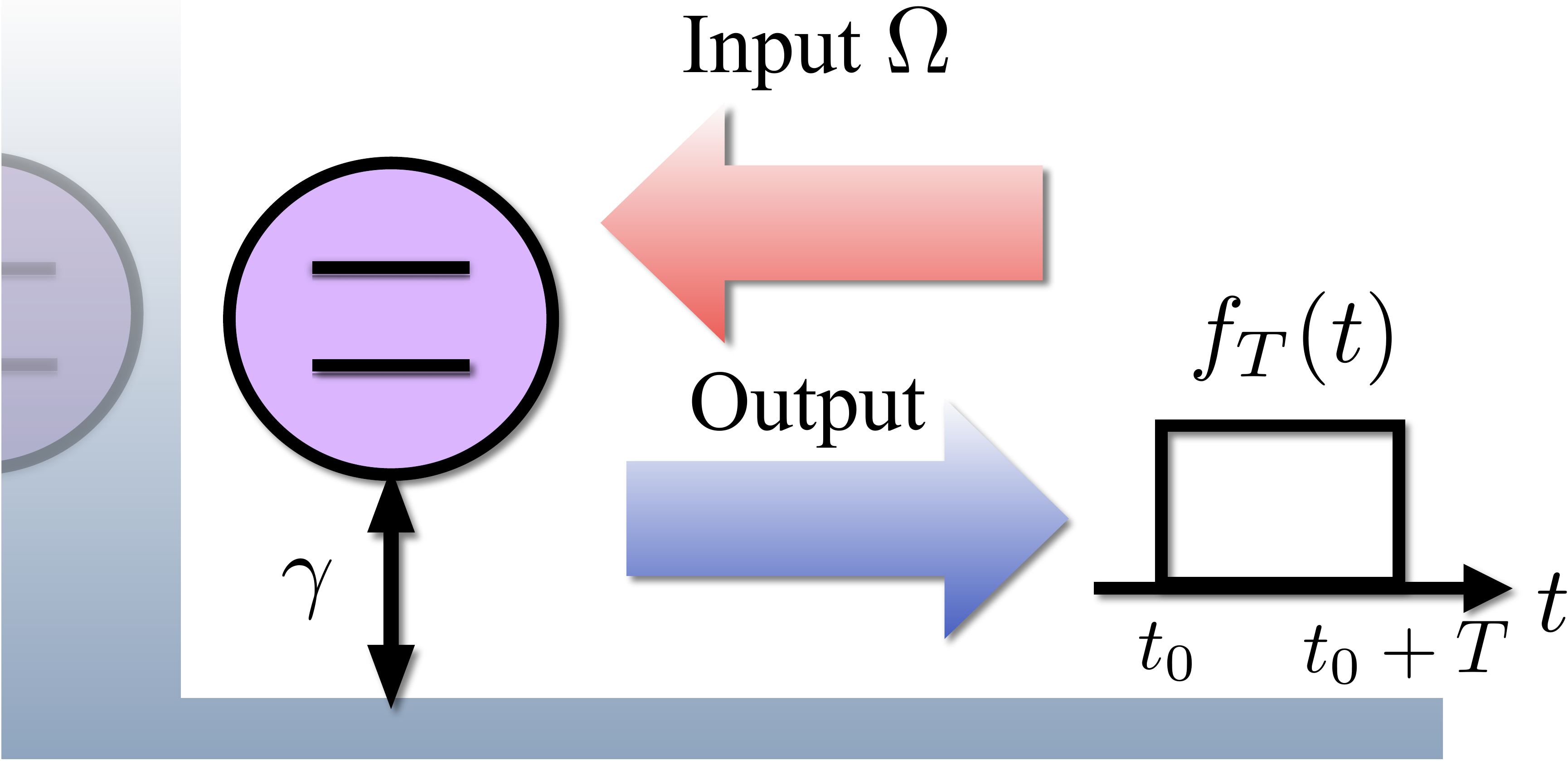}
\end{center}
\caption{(Color online) Two-level system (TLS) in front of a mirror. The TLS
is continuously driven with
a coherent field of amplitude $\Omega$.
In order to reconstruct the state of the output field
it is necessary to filter it in time. Here we use a boxcar
filter of width $T$.
}
\label{fig:setup}
\end{figure}

We use quantum trajectories
~\cite{Wiseman,Brun2002} together with
Maximum Likelihood estimation~\cite{Lvovsky2004, Lvovsky2009}
in order to reconstruct the state and Wigner
function of the scattered field.
We study under which conditions 
this field
is characterized by a negative Wigner function, and furthermore,
quantify its negativity content.

\begin{figure*}[t]
\begin{center}
\includegraphics[width=1.85\columnwidth]{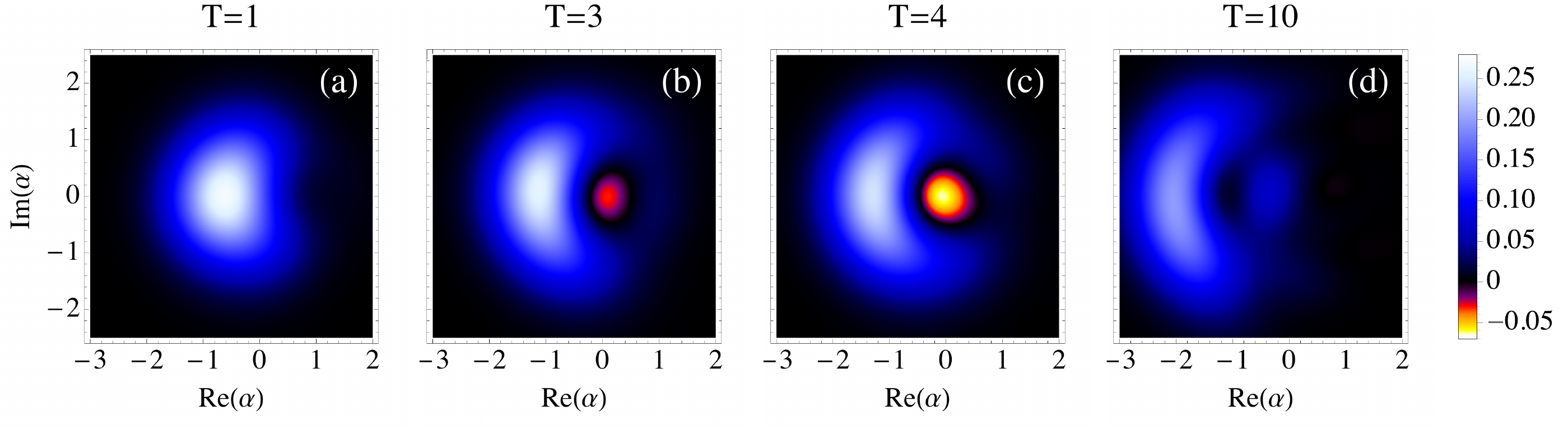}
\end{center}
\caption{(Color online) Contour plots of the Wigner function of
resonance fluorescence of a TLS in front of a mirror at the
incoherent drive point $\Omega^*$.
We set $\gamma = 1$ as the unit.
The filter times correspond to, from (a) to (d): $T=1,\, 3,\, 4,$ and $10$.
At intermediate time scales ((b) and (c))
the Wigner function takes negative values.}
\label{fig:tomo}
\end{figure*}

\paragraph{The setup.---}

In the absence of a mirror, the scattered radiation from a TLS
coupled to a 1D continuum
contains both reflected and transmitted components.
The role of the mirror is to restrict the emission into a single (reflected) component.
The distance between the TLS and the mirror is crucial~\cite{Hoi2015, Pichler2016}.
A non-zero separation
results in time-delay effects which lead to non-Markovian
dynamics~\cite{Pichler2016}.
In this work, we assume
that the TLS-mirror separation is negligible. 

A TLS driven by a resonant
coherent field of amplitude $\Omega$
is described in the rotating frame of the drive by the
Hamiltonian
(we set $\hbar = 1$ hereafter)
\begin{equation}\label{eq-Ham-open}
\hat{H} = - \imi  \sqrt{\gamma} \,\Omega\, ( \hat\sigma_+ - \hat\sigma_-),
\end{equation}
where $\Omega^2$ is the incoming power and
$\hat \sigma_-$ ($\hat \sigma_+$) is the TLS
lowering (raising) operator.
Here $\gamma$ is the coupling strength
to the environment---the electromagnetic field confined in $1$D---
which contains both the incoming coherent drive
and the output field. 

Considering the environment as
a reservoir at zero temperature,
the dissipative dynamics of the TLS is described by the
quantum master equation
\begin{equation}\label{eq-QME-open}
{\rm d}_t \rho = -\imi[\hat H, \rho] + \frac{\gamma}{2} \left( 2 \hat\sigma_- \rho \hat\sigma_+ - \hat\sigma_+ \hat\sigma_- \rho - \rho\, \hat\sigma_+ \hat\sigma_-    \right).
\end{equation}
This setup is sketched in Fig.~\ref{fig:setup}.

The behavior of a TLS in front of a mirror
can be understood by studying the two-time correlations
of the output field~\cite{Gardiner1985}
\begin{equation}\label{eq:in-out}
\hat a_{\rm out}(t) = \Omega + \sqrt{\gamma} \hat \sigma_- (t) .
\end{equation}

From the steady-state of~\eqref{eq-QME-open} (${\rm d}_t \rho = 0$), it is straightforward to
derive the correlation function
\begin{align}\label{eq:two-time}
&\langle \hat a_{\rm out}^\dagger (t) \hat a_{\rm out}(0) \rangle_{\rm ss} =
\left( \Omega - \frac{2 \Omega}{1 + 8 \Omega^2/ \gamma} \right)^2
 \nonumber\\
 &+ \frac{2 \Omega^2}{1 +  8 \Omega^2 / \gamma} \exp \left( - \frac{\gamma t}{2} \right) \nonumber\\
 &+ \lambda_+ \exp \left[ - \frac{\gamma t}{4} \left( 3 + \imi \sqrt{64  \Omega^2/\gamma -1}  \right) \right]
 \nonumber\\
 &+ \lambda_- \exp \left[ - \frac{\gamma t}{4} \left( 3 - \imi \sqrt{64  \Omega^2/\gamma -1}  \right) \right],
\end{align}
where the subscript ss indicates that expectation values are calculated in the steady-state.
Here \mbox{$\lambda_+=\lambda_-^*$} is a function of $\Omega$ and $\gamma$ which we do not write
explicitly for the sake of simplicity (see Supplementary Material~\cite{EPAPS}).
The time-independent part of~\eqref{eq:two-time}
corresponds to
$\langle \hat a^\dagger_{\rm out} \rangle_{\rm ss} \langle \hat a_{\rm out} \rangle_{\rm ss}$, i.e.,
a coherent state.
In both the weak ($\Omega^2/ \gamma \to 0$) and strong
($\Omega^2/ \gamma \to \infty$) driving regimes,
this is the dominant part of the field,
i.e.,
$\langle \hat a^\dagger_{\rm out} (t) \hat a_{\rm out} (0) \rangle_{\rm ss} = \langle \hat a^\dagger_{\rm out} \rangle_{\rm ss} \langle \hat a_{\rm out} \rangle_{\rm ss}$.
For a TLS in front of a mirror
there is a drive strength $\Omega^*$ for which
$\langle \hat a_{\rm out} \rangle_{\rm ss} =  0$~\citep{Hoi2015}.
From~\eqref{eq:two-time}
we find that
 $\Omega^{*2} = \gamma/8 $.
For this drive strength,
the first term on the right-hand-side of
\eqref{eq:two-time} is equal to zero, and we only retain the
time-dependent
terms, meaning the response of the system
is entirely incoherent.
We will refer to $\Omega^*$ as the \textit{incoherent drive point}.
A coherent state is a Gaussian state, and is therefore characterized by a positive Wigner function~\cite{Hudson1974}.
Thus, in order to witness Wigner nonclassicality
we focus on the drive strength
 $\Omega^*$
for which the coherent response
is supressed.

\paragraph{Filtered modes.---}

The Wigner function is defined for a single bosonic mode~\cite{Hillery1984}.
However, the field interacting with the TLS
is a propagating field and
corresponds
to a continuum of modes in time (or frequency).
In order to construct the Wigner function we need to 
pick a single mode out of this continuum.
Following~\cite{Loudon1983},
this can be done using a filter function.
We define the
filtered creation operator
\begin{equation}\label{eq:mode}
\hat A_f^\dagger = \int_0^\infty {\rm d}t\, f(t) \,\hat a_{\rm out}^\dagger (t).
\end{equation}
If the filter function $f$ satisifies the normalization condition
$\int_0^\infty {\rm d}t \vert f(t) \vert^2 = 1$, then the
field $\hat A_f$
obeys the bosonic commutation relation
$[\hat A_f, \hat A_f^\dagger] = 1$.

For simplicity,  we choose to filter the
steady-state emission 
with a
boxcar filter [Cf. Fig.~\ref{fig:setup}]
\begin{equation}\label{eq:boxcar}
f_T(t) = \frac{1}{\sqrt{T}} \left[ \Theta(t-t_0) - \Theta(t-t_0 -T) \right],
\end{equation}
which is a constant function within the time interval
from $t_0$ to $t_0 + T$ and zero elsewhere. Here $t_0$ represents the time
at which the measurement starts and
$\Theta(t)$ is the Heaviside step function.
The \textit{filter time} $T$ defines the only time scale in our problem.
The Fourier transform of
a boxcar filter~\eqref{eq:boxcar} in time is a sinc function in frequency space.
Correspondingly, the filter
imposes an approximate bandwidth of $2/\pi T$---the
width of the sinc central peak.

The filter function 
defines what
is the observed mode upon tomography. 
We have also carried out
the simulations presented in the next section using a Gaussian filter
and the results are essentially the same.

\paragraph{Witnessing Wigner negativity.---}

Homodyne tomography is 
an
experimental technique which allows
to reconstruct the Wigner function of an arbitrary state of light~\cite{Lvovsky2004}.
This relies on homodyne detection~\cite{Lvovsky2009}, i.e., the
measurement of the generalized quadrature operators
\mbox{$\hat a_{\rm out}(t)\, {\rm e}^{-i \phi} + \hat a^\dagger_{\rm out}(t)\, {\rm e}^{+i \phi}$,
with $\phi \in [0, \pi]$}.
The quantum state 
is then
inferred from the
measurement statistics.
Using quantum trajectories~\cite{Wiseman,Brun2002},
we numerically simulate the conditional evolution of the TLS
which results from its emission being subjected to homodyne detection.
The TLS is initialised in its ground state and the measurements
are taken from a large time $t_0$ [Cf. Eq.~\eqref{eq:boxcar}] in which the (unconditional)
master equation~\eqref{eq-QME-open}
has reached the stationary state.
From the quadrature measurement statistics
the state of the field is reconstructed by means of Maximum Likelihood estimation~\cite{Lvovsky2004, Lvovsky2009}.
Technical details can be found in
~\cite{EPAPS, Strandberg2017}.
Knowing the state of the field,
it is straightforward to calculate its Wigner function~\cite{Moyal1949}.
We will focus on the emission from the TLS, that is, we will ignore the reflected drive field in~\eqref{eq:in-out}.
The effect of the latter is to displace
the field emitted by the TLS. This operation
corresponds
to a translation
of the Wigner function in phase space,
which does not affect its negativity~\cite{EPAPS}.

In Fig.~\ref{fig:tomo}
we show the Wigner function
of the output
field
from the TLS in front of a mirror,
driven with strength $\Omega^*$,
for
four different
filter times $T = 1,\,3,\,4$ and $10$
(in units of $\gamma = 1$).
As can be expected based on the discussion after Eq.~\eqref{eq:two-time},
at the incoherent drive point
the output field
is nonclassical as manifested by a negative Wigner
function (Fig.~\ref{fig:neg}(b) and (c)).
This is the main result of this Letter.

\begin{figure}[t]
\begin{center}
\includegraphics[width=0.85\columnwidth]{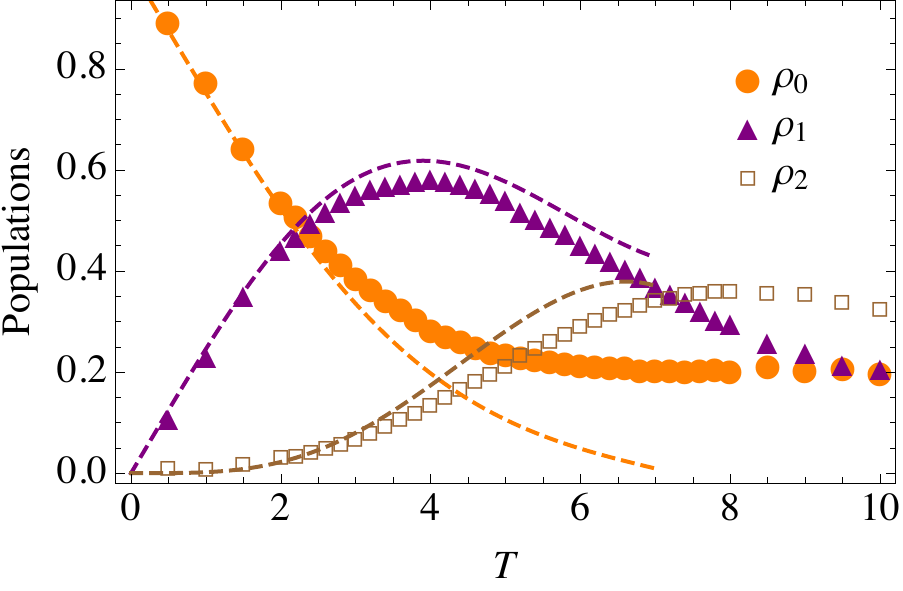}
\end{center}
\caption{(Color online)  Vacuum (filled orange circles), single- (filled purple triangles)
and two-photon
(open brown squares) populations
for the reconstructed state of the output field for a drive strength $\Omega^*$ and as
a function of $T$. Here we set $\gamma =1$ as the unit.
The dashed lines of the same colors correspond to the analytical
solutions of these populations which we include for comparison. 
}
\label{fig:pops}
\end{figure}

\begin{figure}[t]
\begin{center}
\includegraphics[width=0.98\columnwidth]{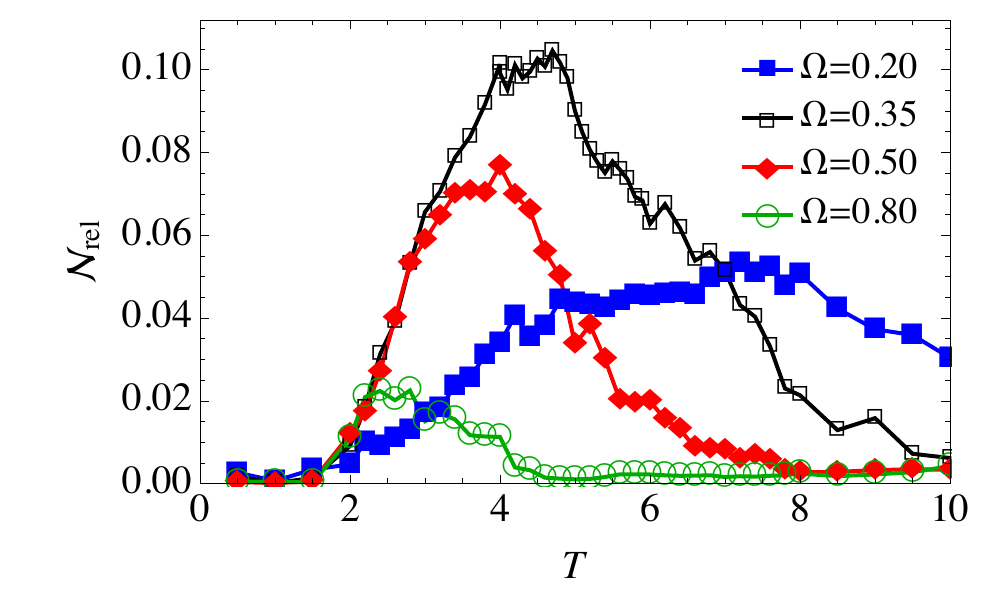}
\end{center}
\caption{
(Color online) Relative negativity $\mathcal{N}_{\rm rel}$
as a function
  of $T$ for $\Omega=0.2$ (filled blue squares),
  $0.35$ (open black squares), $0.5$ (filled red diamonds)
  and $0.8$ (open green circles). Here we set $\gamma =1$ as the unit,
consequently $\Omega^* = 0.35$. 
}
\label{fig:neg}
\end{figure}

In Fig.~\ref{fig:pops}
we show the populations of the photon number states
$\vert 0 \rangle, \,\vert 1 \rangle$ and $\vert 2 \rangle$
as a function of the filter time $T$ at the incoherent drive point.
In this figure we compare the populations for the reconstructed states
from our quantum trajectory simulations with analytical solutions
for the total, \textit{unfiltered}, output field.
The unfiltered field corresponds to the infinite bandwidth limit and
its photon number content can be  calculated analytically
~\cite{Joel2014, Sankar2016}.
For $T \lesssim 2/\gamma$,
there is an agreement
between both solutions.
This is no longer the case for larger values of $T$.
As discussed in the previous section, the boxcar filter
introduces an effective bandwidth of $2/ \pi T$.
Therefore, 
as we approach the infinite bandwidth limit ($T \to 0$)
both solutions agree with each other.
By increasing the filter time, we reduce the effective bandwidth
and we are not able to detect
all of the $\Omega^{* 2} T$ emitted photons.
In addition, for filter times $T \lesssim 2/\gamma$,
two-photon states can
safely be neglected in the output field (they constitute less
than $3 \%$ of the total population).
In the following, we study the \emph{origin} of negativity in a Wigner
function restricted to a two-dimensional Fock space.

The most general state in the space spanned by the
vacuum $\vert 0 \rangle$ and a single-photon $\vert 1 \rangle$ 
is of the form
$\rho = \rho_0 \vert 0 \rangle \langle 0 \vert + \rho_1 \vert 1 \rangle \langle 1 \vert
+ (\rho_{10} \vert 1 \rangle \langle 0 \vert + {\rm h.c.})$, with
the normalization condition \mbox{$\rho_0 + \rho_1 = 1$}. 
The Wigner function
of $\rho$ is~\cite{Moyal1949}
\begin{align}\label{eq:Wigner2}
W_\rho (\alpha) &= \rho_0 W_{\vert 0 \rangle \langle 0 \vert}(\alpha) +
\rho_1 W_{\vert 1 \rangle \langle 1 \vert}(\alpha)  \nonumber\\
& +
\frac{2 \sqrt{2}}{\pi} {\rm e}^{-\vert \alpha \vert^2} {\rm Re} \left[ \rho_{10} \,\alpha \right],
\end{align}
with $W_{\vert n \rangle \langle n \vert}(\alpha) = (-1)^n \exp(-2 \vert \alpha \vert^2)
 L_n[ 4  \vert \alpha \vert^2]/ \pi$ the Wigner function of a Fock state
 $\vert n \rangle$ ($n \geq 0$)~\cite{Gardiner}, and $L_n[x]$ the $n$th order Laguerre polynomial.
In order to quantify the presence of Wigner negativity
we use the
\textit{total integrated
  negativity}~\cite{Kenfack2004,Albarelli2016,Joana2016}, defined as
\begin{equation}\label{eq:int-neg}
\mathcal{N} \equiv \frac{1}{2} \int {\rm d}^2 \alpha \,
\big( \vert W(\alpha) \vert - W(\alpha) \big).
\end{equation}
This is a measure of the
volume of the negative part of the Wigner
function, such that $\mathcal N \geqslant 0$.

In a two-dimensional Fock space,
the presence of negativity ($\mathcal{N} > 0$)
is determined by the populations $\rho_0$ and $\rho_1$.
However, the specific relation between them for a state to be
Wigner-negative is set by the coherences  \mbox{$\rho_{01} = \rho_{10}^*$},
or equivalently,
by the \textit{purity} of the state.
For a pure state
a very small single-photon population ($\rho_1 \sim 0.07$)
is enough for the Wigner function to be negative.
This negativity content is enhanced by increasing
$\rho_1$.
As the purity of the state decreases, the condition for
it to be Wigner-negative is roughly
$\rho_1 \gtrsim \rho_0$. The exact relation
between $\rho_0$ and $\rho_1$ depends on the particular value of the purity.
For an incoherent mixture of $\vert 0 \rangle$ and $\vert 1 \rangle$, i.e.,
\mbox{$\rho_{01} = \rho_{10} = 0$}, the condition becomes
$\rho_1 > \rho_0$.

This analysis is important,
as in general the output state from
an atom in front of a mirror is mixed.
In fact, the output state is pure only in the limit of very weak or strong driving.
Following our discussion on two-time correlations
of the output field, in both regimes the output is a
coherent state. 
Away from
the strong driving regime, 
the purity of the output field decreases with an increasing drive~\cite{EPAPS}.

At the incoherent drive point, from Fig.~\ref{fig:pops}
it follows that near $T = 2/\gamma$, the vacuum and single-photon
contributions to the state become identical.
Following our previous discussion,
 it is around this point where we expect
the Wigner function to become negative as the state is mixed
\cite{EPAPS}.
  To verify this, we calculate the total integrated negativity.
  However, to
  get a feeling for the magnitude of this quantity, instead of
  presenting the negativity~\eqref{eq:int-neg} we show the
  \textit{relative negativity} $\mathcal{N}_{\rm rel}$.
  We define the latter
as the ratio between the total integrated negativity of our state
and the total integrated negativity of a single photon $\mathcal N_{\vert 1 \rangle}\simeq 0.43$: $\mathcal{N}_{\rm rel} \equiv \mathcal N / \mathcal N_{\vert 1 \rangle}$.
In Fig.~\ref{fig:neg}
we show the
relative negativity $\mathcal{N}_{\rm rel}$
as a function of $T$
for different drive strengths.
For $\Omega =\Omega^*$ (open black squares in Fig.~\ref{fig:neg})
we see that there is a correspondence between an
increasing single-photon population
[Cf. Fig.~\ref{fig:pops}]
and the appearance of negativity.
The maximum negativity occurs around $T = 4/\gamma$, and the
corresponding Wigner function is shown in
Fig.~\ref{fig:tomo} (c).
Here, the two-photon population is no longer negligible.
Nevertheless, it is still the dominant single-photon
contribution which renders the state Wigner-negative.
In fact, using the general expression for the Wigner
function of a Fock state (given 
below Eq.~\eqref{eq:Wigner2}), it can be shown
that a small population of $\vert 2 \rangle$
reduces the negativity contribution of $\vert 1 \rangle$
in a mixed state.
Therefore,
it is the
increasing two-photon population which stops the negativity
from growing further.
For longer filter times ($T > 4/\gamma$), 
$\mathcal N_{\rm rel}$ decreases to zero.
Here, higher number states are also involved and the analysis
is not straightforward.
It is worth
to emphasize that we observe Wigner negativity
beyond mixtures of vacuum and single-photons.


We have 
verified that
the largest negativity 
achieved for the output field
occurs at 
$\Omega = \Omega^*$.
In Fig.~\ref{fig:neg}, we compare
the relative negativity for
different drive strengths
in order to show the observed characteristic behavior.
For $\Omega > \Omega^*$ (filled red diamonds and open green circles in Fig.~\ref{fig:neg}),
the output state is also mixed and therefore, using the two-dimensional
Fock space approximation,
the state becomes negative whenever $\rho_1 \gtrsim \rho_0$.
The corresponding populations are shown in~\cite{EPAPS}. 
Also for drive strengths $\Omega > \Omega^*$, the approximation of a two-dimensional
Fock space breaks down at smaller values of $T$ compared to the incoherent drive point.
This means that the single-photon population becomes dominant
at a shorter filter time $T$, which explains the shift in the position
of the maximum $\mathcal N_{\rm rel}$ towards smaller $T$ for stronger drives.
In these cases,
the enhanced presence of higher number states ($\vert n \rangle, \, n \geq 2$)
in the output notably reduces the maximum
negativity achieved.

The case $\Omega < \Omega^*$ (filled blue squares in Fig.~\ref{fig:neg}),
is different
because
the state is almost pure. 
Following the discussion after Eq.~\eqref{eq:int-neg}, a smaller
single-photon population suffices for the state to become negative.
Consequently, the transition from zero to non-zero negativity is less sharp than
for $\Omega \geq \Omega^*$.
For the same reason, negativity is present for a
larger range of values of $T$ compared to stronger drives.

In Ref.~\cite{Schulte2015}, the authors
calculate 
the Wigner function  
of resonance fluorescence analytically
in order to explain the phase-dependent
nature of squeezing.
We believe that their approach, mapping the steady-state of the TLS into the field,
does not correspond to steady-state emission.
It is however
correct if the drive 
is switched off once the
steady-state is reached and the emission from the TLS is homodyned with a
filter matching its decay in time.

The setup presented in this Letter
is already experimentally feasible. Most notable realizations are
found in superconducting
circuits and quantum dots.
In superconducting circuits, a measurement scheme
closely resembling 
homodyne detection
corresponds to phase sensitive amplification \cite{Mallet2011}.
Alternatively, there are other schemes for characterizing
propagating fields~\cite{Menzel2010, daSilva2010, Eichler2011, Menzel2012, Fedorov2016}.
Quantum state tomography for the scattering of coherent states
on an array of quantum dots has already been reported
\cite{Boehm2017}.

\paragraph{Conclusions.---}
We have calculated the Wigner function of
$1$D
resonance fluorescence from a TLS
in front of a mirror.
We use the mirror to exactly cancel the coherent
emission from the TLS.
We have shown that at the incoherent drive
point
the Wigner function achieves its maximum
negativity content.

Protocols have been demonstrated where, e.g., the cubic phase
gate, which allows to promote the Gaussian set of gates to a universal set~\cite{Gu2009},
can be obtained by using input non-Gaussian ancillary states together
with Gaussian operations and
measurements~\cite{Arzani2017,Albarelli2018}. Therefore, we have verified that this
simple setup suffices
to generate the class of states necessary for universal
quantum computing beyond the scope of classical computers.

\paragraph{Acknowledgments.---}

The authors would like to thank
Giulia Ferrini,
Sankar Raman Sathyamoorthy,
Adam Miranowicz,
Sahin K. \"Ozdemir and Steven Girvin for valuable discussions.
FQ and GJ acknowledge the support from the Knut and Alice Wallenberg Foundation.
IS acknowledges the support from Chalmers Excellence Initiative Nano.

\bibliographystyle{apsrev4-1}
\bibliography{wigner}

\begin{thebibliography}{55}%
\makeatletter
\providecommand \@ifxundefined [1]{%
 \@ifx{#1\undefined}
}%
\providecommand \@ifnum [1]{%
 \ifnum #1\expandafter \@firstoftwo
 \else \expandafter \@secondoftwo
 \fi
}%
\providecommand \@ifx [1]{%
 \ifx #1\expandafter \@firstoftwo
 \else \expandafter \@secondoftwo
 \fi
}%
\providecommand \natexlab [1]{#1}%
\providecommand \enquote  [1]{``#1''}%
\providecommand \bibnamefont  [1]{#1}%
\providecommand \bibfnamefont [1]{#1}%
\providecommand \citenamefont [1]{#1}%
\providecommand \href@noop [0]{\@secondoftwo}%
\providecommand \href [0]{\begingroup \@sanitize@url \@href}%
\providecommand \@href[1]{\@@startlink{#1}\@@href}%
\providecommand \@@href[1]{\endgroup#1\@@endlink}%
\providecommand \@sanitize@url [0]{\catcode `\\12\catcode `\$12\catcode
  `\&12\catcode `\#12\catcode `\^12\catcode `\_12\catcode `\%12\relax}%
\providecommand \@@startlink[1]{}%
\providecommand \@@endlink[0]{}%
\providecommand \url  [0]{\begingroup\@sanitize@url \@url }%
\providecommand \@url [1]{\endgroup\@href {#1}{\urlprefix }}%
\providecommand \urlprefix  [0]{URL }%
\providecommand \Eprint [0]{\href }%
\providecommand \doibase [0]{http://dx.doi.org/}%
\providecommand \selectlanguage [0]{\@gobble}%
\providecommand \bibinfo  [0]{\@secondoftwo}%
\providecommand \bibfield  [0]{\@secondoftwo}%
\providecommand \translation [1]{[#1]}%
\providecommand \BibitemOpen [0]{}%
\providecommand \bibitemStop [0]{}%
\providecommand \bibitemNoStop [0]{.\EOS\space}%
\providecommand \EOS [0]{\spacefactor3000\relax}%
\providecommand \BibitemShut  [1]{\csname bibitem#1\endcsname}%
\let\auto@bib@innerbib\@empty
\bibitem [{\citenamefont {Walls}\ and\ \citenamefont {Milburn}(2008)}]{Walls}%
  \BibitemOpen
  \bibfield  {author} {\bibinfo {author} {\bibfnamefont {D.}~\bibnamefont
  {Walls}}\ and\ \bibinfo {author} {\bibfnamefont {G.}~\bibnamefont
  {Milburn}},\ }\href@noop {} {\emph {\bibinfo {title} {Quantum Optics}}}\
  (\bibinfo  {publisher} {Springer Berlin Heidelberg},\ \bibinfo {year}
  {2008})\BibitemShut {NoStop}%
\bibitem [{\citenamefont {Kimble}\ \emph {et~al.}(1977)\citenamefont {Kimble},
  \citenamefont {Dagenais},\ and\ \citenamefont {Mandel}}]{Kimble1977}%
  \BibitemOpen
  \bibfield  {author} {\bibinfo {author} {\bibfnamefont {H.~J.}\ \bibnamefont
  {Kimble}}, \bibinfo {author} {\bibfnamefont {M.}~\bibnamefont {Dagenais}}, \
  and\ \bibinfo {author} {\bibfnamefont {L.}~\bibnamefont {Mandel}},\ }\href
  {\doibase 10.1103/PhysRevLett.39.691} {\bibfield  {journal} {\bibinfo
  {journal} {Phys. Rev. Lett.}\ }\textbf {\bibinfo {volume} {39}},\ \bibinfo
  {pages} {691} (\bibinfo {year} {1977})}\BibitemShut {NoStop}%
\bibitem [{\citenamefont {Walls}\ and\ \citenamefont
  {Zoller}(1981)}]{Walls1981}%
  \BibitemOpen
  \bibfield  {author} {\bibinfo {author} {\bibfnamefont {D.~F.}\ \bibnamefont
  {Walls}}\ and\ \bibinfo {author} {\bibfnamefont {P.}~\bibnamefont {Zoller}},\
  }\href {\doibase 10.1103/PhysRevLett.47.709} {\bibfield  {journal} {\bibinfo
  {journal} {Phys. Rev. Lett.}\ }\textbf {\bibinfo {volume} {47}},\ \bibinfo
  {pages} {709} (\bibinfo {year} {1981})}\BibitemShut {NoStop}%
\bibitem [{\citenamefont {Mollow}(1969)}]{Mollow1969}%
  \BibitemOpen
  \bibfield  {author} {\bibinfo {author} {\bibfnamefont {B.~R.}\ \bibnamefont
  {Mollow}},\ }\href {\doibase 10.1103/PhysRev.188.1969} {\bibfield  {journal}
  {\bibinfo  {journal} {Phys. Rev.}\ }\textbf {\bibinfo {volume} {188}},\
  \bibinfo {pages} {1969} (\bibinfo {year} {1969})}\BibitemShut {NoStop}%
\bibitem [{\citenamefont {Wigner}(1932)}]{Wigner1932}%
  \BibitemOpen
  \bibfield  {author} {\bibinfo {author} {\bibfnamefont {E.}~\bibnamefont
  {Wigner}},\ }\href {\doibase 10.1103/PhysRev.40.749} {\bibfield  {journal}
  {\bibinfo  {journal} {Phys. Rev.}\ }\textbf {\bibinfo {volume} {40}},\
  \bibinfo {pages} {749} (\bibinfo {year} {1932})}\BibitemShut {NoStop}%
\bibitem [{\citenamefont {Hillery}\ \emph {et~al.}(1984)\citenamefont
  {Hillery}, \citenamefont {O'Connell}, \citenamefont {Scully},\ and\
  \citenamefont {Wigner}}]{Hillery1984}%
  \BibitemOpen
  \bibfield  {author} {\bibinfo {author} {\bibfnamefont {M.}~\bibnamefont
  {Hillery}}, \bibinfo {author} {\bibfnamefont {R.}~\bibnamefont {O'Connell}},
  \bibinfo {author} {\bibfnamefont {M.}~\bibnamefont {Scully}}, \ and\ \bibinfo
  {author} {\bibfnamefont {E.}~\bibnamefont {Wigner}},\ }\href {\doibase
  https://doi.org/10.1016/0370-1573(84)90160-1} {\bibfield  {journal} {\bibinfo
   {journal} {Physics Reports}\ }\textbf {\bibinfo {volume} {106}},\ \bibinfo
  {pages} {121 } (\bibinfo {year} {1984})}\BibitemShut {NoStop}%
\bibitem [{\citenamefont {Gardiner}\ and\ \citenamefont
  {Zoller}(2004)}]{Gardiner}%
  \BibitemOpen
  \bibfield  {author} {\bibinfo {author} {\bibfnamefont {C.}~\bibnamefont
  {Gardiner}}\ and\ \bibinfo {author} {\bibfnamefont {P.}~\bibnamefont
  {Zoller}},\ }\href@noop {} {\emph {\bibinfo {title} {Quantum Noise}}},\
  Springer Series in Synergetics\ (\bibinfo  {publisher} {Springer},\ \bibinfo
  {year} {2004})\BibitemShut {NoStop}%
\bibitem [{\citenamefont {Lloyd}\ and\ \citenamefont
  {Braunstein}(1999)}]{Braunstein1999}%
  \BibitemOpen
  \bibfield  {author} {\bibinfo {author} {\bibfnamefont {S.}~\bibnamefont
  {Lloyd}}\ and\ \bibinfo {author} {\bibfnamefont {S.~L.}\ \bibnamefont
  {Braunstein}},\ }\href {\doibase 10.1103/PhysRevLett.82.1784} {\bibfield
  {journal} {\bibinfo  {journal} {Phys. Rev. Lett.}\ }\textbf {\bibinfo
  {volume} {82}},\ \bibinfo {pages} {1784} (\bibinfo {year}
  {1999})}\BibitemShut {NoStop}%
\bibitem [{\citenamefont {Gu}\ \emph {et~al.}(2009)\citenamefont {Gu},
  \citenamefont {Weedbrook}, \citenamefont {Menicucci}, \citenamefont {Ralph},\
  and\ \citenamefont {van Loock}}]{Gu2009}%
  \BibitemOpen
  \bibfield  {author} {\bibinfo {author} {\bibfnamefont {M.}~\bibnamefont
  {Gu}}, \bibinfo {author} {\bibfnamefont {C.}~\bibnamefont {Weedbrook}},
  \bibinfo {author} {\bibfnamefont {N.~C.}\ \bibnamefont {Menicucci}}, \bibinfo
  {author} {\bibfnamefont {T.~C.}\ \bibnamefont {Ralph}}, \ and\ \bibinfo
  {author} {\bibfnamefont {P.}~\bibnamefont {van Loock}},\ }\href {\doibase
  10.1103/PhysRevA.79.062318} {\bibfield  {journal} {\bibinfo  {journal} {Phys.
  Rev. A}\ }\textbf {\bibinfo {volume} {79}},\ \bibinfo {pages} {062318}
  (\bibinfo {year} {2009})}\BibitemShut {NoStop}%
\bibitem [{\citenamefont {Mari}\ and\ \citenamefont {Eisert}(2012)}]{Mari2012}%
  \BibitemOpen
  \bibfield  {author} {\bibinfo {author} {\bibfnamefont {A.}~\bibnamefont
  {Mari}}\ and\ \bibinfo {author} {\bibfnamefont {J.}~\bibnamefont {Eisert}},\
  }\href {\doibase 10.1103/PhysRevLett.109.230503} {\bibfield  {journal}
  {\bibinfo  {journal} {Phys. Rev. Lett.}\ }\textbf {\bibinfo {volume} {109}},\
  \bibinfo {pages} {230503} (\bibinfo {year} {2012})}\BibitemShut {NoStop}%
\bibitem [{\citenamefont {Veitch}\ \emph {et~al.}(2013)\citenamefont {Veitch},
  \citenamefont {Wiebe}, \citenamefont {Ferrie},\ and\ \citenamefont
  {Emerson}}]{Veitch2013}%
  \BibitemOpen
  \bibfield  {author} {\bibinfo {author} {\bibfnamefont {V.}~\bibnamefont
  {Veitch}}, \bibinfo {author} {\bibfnamefont {N.}~\bibnamefont {Wiebe}},
  \bibinfo {author} {\bibfnamefont {C.}~\bibnamefont {Ferrie}}, \ and\ \bibinfo
  {author} {\bibfnamefont {J.}~\bibnamefont {Emerson}},\ }\href
  {http://stacks.iop.org/1367-2630/15/i=1/a=013037} {\bibfield  {journal}
  {\bibinfo  {journal} {New Journal of Physics}\ }\textbf {\bibinfo {volume}
  {15}},\ \bibinfo {pages} {013037} (\bibinfo {year} {2013})}\BibitemShut
  {NoStop}%
\bibitem [{\citenamefont {Rahimi-Keshari}\ \emph {et~al.}(2016)\citenamefont
  {Rahimi-Keshari}, \citenamefont {Ralph},\ and\ \citenamefont
  {Caves}}]{Rahimi2016}%
  \BibitemOpen
  \bibfield  {author} {\bibinfo {author} {\bibfnamefont {S.}~\bibnamefont
  {Rahimi-Keshari}}, \bibinfo {author} {\bibfnamefont {T.~C.}\ \bibnamefont
  {Ralph}}, \ and\ \bibinfo {author} {\bibfnamefont {C.~M.}\ \bibnamefont
  {Caves}},\ }\href {\doibase 10.1103/PhysRevX.6.021039} {\bibfield  {journal}
  {\bibinfo  {journal} {Phys. Rev. X}\ }\textbf {\bibinfo {volume} {6}},\
  \bibinfo {pages} {021039} (\bibinfo {year} {2016})}\BibitemShut {NoStop}%
\bibitem [{\citenamefont {{Chang}}\ \emph {et~al.}(2007)\citenamefont
  {{Chang}}, \citenamefont {{S{\o}rensen}}, \citenamefont {{Demler}},\ and\
  \citenamefont {{Lukin}}}]{Chang2007}%
  \BibitemOpen
  \bibfield  {author} {\bibinfo {author} {\bibfnamefont {D.~E.}\ \bibnamefont
  {{Chang}}}, \bibinfo {author} {\bibfnamefont {A.~S.}\ \bibnamefont
  {{S{\o}rensen}}}, \bibinfo {author} {\bibfnamefont {E.~A.}\ \bibnamefont
  {{Demler}}}, \ and\ \bibinfo {author} {\bibfnamefont {M.~D.}\ \bibnamefont
  {{Lukin}}},\ }\href {\doibase 10.1038/nphys708} {\bibfield  {journal}
  {\bibinfo  {journal} {Nature Physics}\ }\textbf {\bibinfo {volume} {3}},\
  \bibinfo {pages} {807} (\bibinfo {year} {2007})}\BibitemShut {NoStop}%
\bibitem [{\citenamefont {Muller}\ \emph {et~al.}(2007)\citenamefont {Muller},
  \citenamefont {Flagg}, \citenamefont {Bianucci}, \citenamefont {Wang},
  \citenamefont {Deppe}, \citenamefont {Ma}, \citenamefont {Zhang},
  \citenamefont {Salamo}, \citenamefont {Xiao},\ and\ \citenamefont
  {Shih}}]{Muller2007}%
  \BibitemOpen
  \bibfield  {author} {\bibinfo {author} {\bibfnamefont {A.}~\bibnamefont
  {Muller}}, \bibinfo {author} {\bibfnamefont {E.~B.}\ \bibnamefont {Flagg}},
  \bibinfo {author} {\bibfnamefont {P.}~\bibnamefont {Bianucci}}, \bibinfo
  {author} {\bibfnamefont {X.~Y.}\ \bibnamefont {Wang}}, \bibinfo {author}
  {\bibfnamefont {D.~G.}\ \bibnamefont {Deppe}}, \bibinfo {author}
  {\bibfnamefont {W.}~\bibnamefont {Ma}}, \bibinfo {author} {\bibfnamefont
  {J.}~\bibnamefont {Zhang}}, \bibinfo {author} {\bibfnamefont {G.~J.}\
  \bibnamefont {Salamo}}, \bibinfo {author} {\bibfnamefont {M.}~\bibnamefont
  {Xiao}}, \ and\ \bibinfo {author} {\bibfnamefont {C.~K.}\ \bibnamefont
  {Shih}},\ }\href {\doibase 10.1103/PhysRevLett.99.187402} {\bibfield
  {journal} {\bibinfo  {journal} {Phys. Rev. Lett.}\ }\textbf {\bibinfo
  {volume} {99}},\ \bibinfo {pages} {187402} (\bibinfo {year}
  {2007})}\BibitemShut {NoStop}%
\bibitem [{\citenamefont {{Astafiev}}\ \emph {et~al.}(2010)\citenamefont
  {{Astafiev}}, \citenamefont {{Zagoskin}}, \citenamefont {{Abdumalikov}},
  \citenamefont {{Pashkin}}, \citenamefont {{Yamamoto}}, \citenamefont
  {{Inomata}}, \citenamefont {{Nakamura}},\ and\ \citenamefont
  {{Tsai}}}]{Astafiev2010}%
  \BibitemOpen
  \bibfield  {author} {\bibinfo {author} {\bibfnamefont {O.}~\bibnamefont
  {{Astafiev}}}, \bibinfo {author} {\bibfnamefont {A.~M.}\ \bibnamefont
  {{Zagoskin}}}, \bibinfo {author} {\bibfnamefont {A.~A.}\ \bibnamefont
  {{Abdumalikov}}}, \bibinfo {author} {\bibfnamefont {Y.~A.}\ \bibnamefont
  {{Pashkin}}}, \bibinfo {author} {\bibfnamefont {T.}~\bibnamefont
  {{Yamamoto}}}, \bibinfo {author} {\bibfnamefont {K.}~\bibnamefont
  {{Inomata}}}, \bibinfo {author} {\bibfnamefont {Y.}~\bibnamefont
  {{Nakamura}}}, \ and\ \bibinfo {author} {\bibfnamefont {J.~S.}\ \bibnamefont
  {{Tsai}}},\ }\href {\doibase 10.1126/science.1181918} {\bibfield  {journal}
  {\bibinfo  {journal} {Science}\ }\textbf {\bibinfo {volume} {327}},\ \bibinfo
  {pages} {840} (\bibinfo {year} {2010})}\BibitemShut {NoStop}%
\bibitem [{\citenamefont {Peropadre}\ \emph {et~al.}(2013)\citenamefont
  {Peropadre}, \citenamefont {Lindkvist}, \citenamefont {Hoi}, \citenamefont
  {Wilson}, \citenamefont {Garcia-Ripoll}, \citenamefont {Delsing},\ and\
  \citenamefont {Johansson}}]{Peropadre2013}%
  \BibitemOpen
  \bibfield  {author} {\bibinfo {author} {\bibfnamefont {B.}~\bibnamefont
  {Peropadre}}, \bibinfo {author} {\bibfnamefont {J.}~\bibnamefont
  {Lindkvist}}, \bibinfo {author} {\bibfnamefont {I.-C.}\ \bibnamefont {Hoi}},
  \bibinfo {author} {\bibfnamefont {C.~M.}\ \bibnamefont {Wilson}}, \bibinfo
  {author} {\bibfnamefont {J.~J.}\ \bibnamefont {Garcia-Ripoll}}, \bibinfo
  {author} {\bibfnamefont {P.}~\bibnamefont {Delsing}}, \ and\ \bibinfo
  {author} {\bibfnamefont {G.}~\bibnamefont {Johansson}},\ }\href
  {http://stacks.iop.org/1367-2630/15/i=3/a=035009} {\bibfield  {journal}
  {\bibinfo  {journal} {New Journal of Physics}\ }\textbf {\bibinfo {volume}
  {15}},\ \bibinfo {pages} {035009} (\bibinfo {year} {2013})}\BibitemShut
  {NoStop}%
\bibitem [{\citenamefont {Hoi}\ \emph {et~al.}(2011)\citenamefont {Hoi},
  \citenamefont {Wilson}, \citenamefont {Johansson}, \citenamefont {Palomaki},
  \citenamefont {Peropadre},\ and\ \citenamefont {Delsing}}]{Hoi2011}%
  \BibitemOpen
  \bibfield  {author} {\bibinfo {author} {\bibfnamefont {I.-C.}\ \bibnamefont
  {Hoi}}, \bibinfo {author} {\bibfnamefont {C.~M.}\ \bibnamefont {Wilson}},
  \bibinfo {author} {\bibfnamefont {G.}~\bibnamefont {Johansson}}, \bibinfo
  {author} {\bibfnamefont {T.}~\bibnamefont {Palomaki}}, \bibinfo {author}
  {\bibfnamefont {B.}~\bibnamefont {Peropadre}}, \ and\ \bibinfo {author}
  {\bibfnamefont {P.}~\bibnamefont {Delsing}},\ }\href {\doibase
  10.1103/PhysRevLett.107.073601} {\bibfield  {journal} {\bibinfo  {journal}
  {Phys. Rev. Lett.}\ }\textbf {\bibinfo {volume} {107}},\ \bibinfo {pages}
  {073601} (\bibinfo {year} {2011})}\BibitemShut {NoStop}%
\bibitem [{\citenamefont {{Schulte}}\ \emph {et~al.}(2015)\citenamefont
  {{Schulte}}, \citenamefont {{Hansom}}, \citenamefont {{Jones}}, \citenamefont
  {{Matthiesen}}, \citenamefont {{Le Gall}},\ and\ \citenamefont
  {{Atat{\"u}re}}}]{Schulte2015}%
  \BibitemOpen
  \bibfield  {author} {\bibinfo {author} {\bibfnamefont {C.~H.~H.}\
  \bibnamefont {{Schulte}}}, \bibinfo {author} {\bibfnamefont {J.}~\bibnamefont
  {{Hansom}}}, \bibinfo {author} {\bibfnamefont {A.~E.}\ \bibnamefont
  {{Jones}}}, \bibinfo {author} {\bibfnamefont {C.}~\bibnamefont
  {{Matthiesen}}}, \bibinfo {author} {\bibfnamefont {C.}~\bibnamefont {{Le
  Gall}}}, \ and\ \bibinfo {author} {\bibfnamefont {M.}~\bibnamefont
  {{Atat{\"u}re}}},\ }\href {\doibase 10.1038/nature14868} {\bibfield
  {journal} {\bibinfo  {journal} {\nat}\ }\textbf {\bibinfo {volume} {525}},\
  \bibinfo {pages} {222} (\bibinfo {year} {2015})}\BibitemShut {NoStop}%
\bibitem [{\citenamefont {Gu}\ \emph {et~al.}(2017)\citenamefont {Gu},
  \citenamefont {Kockum}, \citenamefont {Miranowicz}, \citenamefont {xi~Liu},\
  and\ \citenamefont {Nori}}]{Gu2017}%
  \BibitemOpen
  \bibfield  {author} {\bibinfo {author} {\bibfnamefont {X.}~\bibnamefont
  {Gu}}, \bibinfo {author} {\bibfnamefont {A.~F.}\ \bibnamefont {Kockum}},
  \bibinfo {author} {\bibfnamefont {A.}~\bibnamefont {Miranowicz}}, \bibinfo
  {author} {\bibfnamefont {Y.}~\bibnamefont {xi~Liu}}, \ and\ \bibinfo {author}
  {\bibfnamefont {F.}~\bibnamefont {Nori}},\ }\href {\doibase
  https://doi.org/10.1016/j.physrep.2017.10.002} {\bibfield  {journal}
  {\bibinfo  {journal} {Physics Reports}\ }\textbf {\bibinfo {volume}
  {718-719}},\ \bibinfo {pages} {1 } (\bibinfo {year} {2017})},\ \bibinfo
  {note} {microwave photonics with superconducting quantum
  circuits}\BibitemShut {NoStop}%
\bibitem [{\citenamefont {Koshino}\ and\ \citenamefont
  {Nakamura}(2012)}]{Koshino2012}%
  \BibitemOpen
  \bibfield  {author} {\bibinfo {author} {\bibfnamefont {K.}~\bibnamefont
  {Koshino}}\ and\ \bibinfo {author} {\bibfnamefont {Y.}~\bibnamefont
  {Nakamura}},\ }\href {http://stacks.iop.org/1367-2630/14/i=4/a=043005}
  {\bibfield  {journal} {\bibinfo  {journal} {New Journal of Physics}\ }\textbf
  {\bibinfo {volume} {14}},\ \bibinfo {pages} {043005} (\bibinfo {year}
  {2012})}\BibitemShut {NoStop}%
\bibitem [{\citenamefont {{Hoi}}\ \emph {et~al.}(2015)\citenamefont {{Hoi}},
  \citenamefont {{Kockum}}, \citenamefont {{Tornberg}}, \citenamefont
  {{Pourkabirian}}, \citenamefont {{Johansson}}, \citenamefont {{Delsing}},\
  and\ \citenamefont {{Wilson}}}]{Hoi2015}%
  \BibitemOpen
  \bibfield  {author} {\bibinfo {author} {\bibfnamefont {I.-C.}\ \bibnamefont
  {{Hoi}}}, \bibinfo {author} {\bibfnamefont {A.~F.}\ \bibnamefont {{Kockum}}},
  \bibinfo {author} {\bibfnamefont {L.}~\bibnamefont {{Tornberg}}}, \bibinfo
  {author} {\bibfnamefont {A.}~\bibnamefont {{Pourkabirian}}}, \bibinfo
  {author} {\bibfnamefont {G.}~\bibnamefont {{Johansson}}}, \bibinfo {author}
  {\bibfnamefont {P.}~\bibnamefont {{Delsing}}}, \ and\ \bibinfo {author}
  {\bibfnamefont {C.~M.}\ \bibnamefont {{Wilson}}},\ }\href@noop {} {\bibfield
  {journal} {\bibinfo  {journal} {Nature Physics}\ }\textbf {\bibinfo {volume}
  {11}},\ \bibinfo {pages} {1045} (\bibinfo {year} {2015})}\BibitemShut
  {NoStop}%
\bibitem [{\citenamefont {Eichler}\ \emph {et~al.}(2011)\citenamefont
  {Eichler}, \citenamefont {Bozyigit}, \citenamefont {Lang}, \citenamefont
  {Baur}, \citenamefont {Steffen}, \citenamefont {Fink}, \citenamefont
  {Filipp},\ and\ \citenamefont {Wallraff}}]{Eichler2011}%
  \BibitemOpen
  \bibfield  {author} {\bibinfo {author} {\bibfnamefont {C.}~\bibnamefont
  {Eichler}}, \bibinfo {author} {\bibfnamefont {D.}~\bibnamefont {Bozyigit}},
  \bibinfo {author} {\bibfnamefont {C.}~\bibnamefont {Lang}}, \bibinfo {author}
  {\bibfnamefont {M.}~\bibnamefont {Baur}}, \bibinfo {author} {\bibfnamefont
  {L.}~\bibnamefont {Steffen}}, \bibinfo {author} {\bibfnamefont {J.~M.}\
  \bibnamefont {Fink}}, \bibinfo {author} {\bibfnamefont {S.}~\bibnamefont
  {Filipp}}, \ and\ \bibinfo {author} {\bibfnamefont {A.}~\bibnamefont
  {Wallraff}},\ }\href@noop {} {\bibfield  {journal} {\bibinfo  {journal}
  {Phys. Rev. Lett.}\ }\textbf {\bibinfo {volume} {107}},\ \bibinfo {pages}
  {113601} (\bibinfo {year} {2011})}\BibitemShut {NoStop}%
\bibitem [{\citenamefont {Yin}\ \emph {et~al.}(2013)\citenamefont {Yin},
  \citenamefont {Chen}, \citenamefont {Sank}, \citenamefont {O'Malley},
  \citenamefont {White}, \citenamefont {Barends}, \citenamefont {Kelly},
  \citenamefont {Lucero}, \citenamefont {Mariantoni}, \citenamefont {Megrant},
  \citenamefont {Neill}, \citenamefont {Vainsencher}, \citenamefont {Wenner},
  \citenamefont {Korotkov}, \citenamefont {Cleland},\ and\ \citenamefont
  {Martinis}}]{Yin2013}%
  \BibitemOpen
  \bibfield  {author} {\bibinfo {author} {\bibfnamefont {Y.}~\bibnamefont
  {Yin}}, \bibinfo {author} {\bibfnamefont {Y.}~\bibnamefont {Chen}}, \bibinfo
  {author} {\bibfnamefont {D.}~\bibnamefont {Sank}}, \bibinfo {author}
  {\bibfnamefont {P.~J.~J.}\ \bibnamefont {O'Malley}}, \bibinfo {author}
  {\bibfnamefont {T.~C.}\ \bibnamefont {White}}, \bibinfo {author}
  {\bibfnamefont {R.}~\bibnamefont {Barends}}, \bibinfo {author} {\bibfnamefont
  {J.}~\bibnamefont {Kelly}}, \bibinfo {author} {\bibfnamefont
  {E.}~\bibnamefont {Lucero}}, \bibinfo {author} {\bibfnamefont
  {M.}~\bibnamefont {Mariantoni}}, \bibinfo {author} {\bibfnamefont
  {A.}~\bibnamefont {Megrant}}, \bibinfo {author} {\bibfnamefont
  {C.}~\bibnamefont {Neill}}, \bibinfo {author} {\bibfnamefont
  {A.}~\bibnamefont {Vainsencher}}, \bibinfo {author} {\bibfnamefont
  {J.}~\bibnamefont {Wenner}}, \bibinfo {author} {\bibfnamefont {A.~N.}\
  \bibnamefont {Korotkov}}, \bibinfo {author} {\bibfnamefont {A.~N.}\
  \bibnamefont {Cleland}}, \ and\ \bibinfo {author} {\bibfnamefont {J.~M.}\
  \bibnamefont {Martinis}},\ }\href {\doibase 10.1103/PhysRevLett.110.107001}
  {\bibfield  {journal} {\bibinfo  {journal} {Phys. Rev. Lett.}\ }\textbf
  {\bibinfo {volume} {110}},\ \bibinfo {pages} {107001} (\bibinfo {year}
  {2013})}\BibitemShut {NoStop}%
\bibitem [{\citenamefont {Sathyamoorthy}\ \emph {et~al.}(2016)\citenamefont
  {Sathyamoorthy}, \citenamefont {Bengtsson}, \citenamefont {Bens},
  \citenamefont {Simoen}, \citenamefont {Delsing},\ and\ \citenamefont
  {Johansson}}]{Sankar2016}%
  \BibitemOpen
  \bibfield  {author} {\bibinfo {author} {\bibfnamefont {S.~R.}\ \bibnamefont
  {Sathyamoorthy}}, \bibinfo {author} {\bibfnamefont {A.}~\bibnamefont
  {Bengtsson}}, \bibinfo {author} {\bibfnamefont {S.}~\bibnamefont {Bens}},
  \bibinfo {author} {\bibfnamefont {M.}~\bibnamefont {Simoen}}, \bibinfo
  {author} {\bibfnamefont {P.}~\bibnamefont {Delsing}}, \ and\ \bibinfo
  {author} {\bibfnamefont {G.}~\bibnamefont {Johansson}},\ }\href@noop {}
  {\bibfield  {journal} {\bibinfo  {journal} {Phys. Rev. A}\ }\textbf {\bibinfo
  {volume} {93}},\ \bibinfo {pages} {063823} (\bibinfo {year}
  {2016})}\BibitemShut {NoStop}%
\bibitem [{\citenamefont {Forn-D\'{\i}az}\ \emph {et~al.}(2017)\citenamefont
  {Forn-D\'{\i}az}, \citenamefont {Warren}, \citenamefont {Chang},
  \citenamefont {Vadiraj},\ and\ \citenamefont {Wilson}}]{Forn-Diaz2017}%
  \BibitemOpen
  \bibfield  {author} {\bibinfo {author} {\bibfnamefont {P.}~\bibnamefont
  {Forn-D\'{\i}az}}, \bibinfo {author} {\bibfnamefont {C.~W.}\ \bibnamefont
  {Warren}}, \bibinfo {author} {\bibfnamefont {C.~W.~S.}\ \bibnamefont
  {Chang}}, \bibinfo {author} {\bibfnamefont {A.~M.}\ \bibnamefont {Vadiraj}},
  \ and\ \bibinfo {author} {\bibfnamefont {C.~M.}\ \bibnamefont {Wilson}},\
  }\href {\doibase 10.1103/PhysRevApplied.8.054015} {\bibfield  {journal}
  {\bibinfo  {journal} {Phys. Rev. Applied}\ }\textbf {\bibinfo {volume} {8}},\
  \bibinfo {pages} {054015} (\bibinfo {year} {2017})}\BibitemShut {NoStop}%
\bibitem [{\citenamefont {Matthiesen}\ \emph {et~al.}(2012)\citenamefont
  {Matthiesen}, \citenamefont {Vamivakas},\ and\ \citenamefont
  {Atat\"ure}}]{Matthiesen2012}%
  \BibitemOpen
  \bibfield  {author} {\bibinfo {author} {\bibfnamefont {C.}~\bibnamefont
  {Matthiesen}}, \bibinfo {author} {\bibfnamefont {A.~N.}\ \bibnamefont
  {Vamivakas}}, \ and\ \bibinfo {author} {\bibfnamefont {M.}~\bibnamefont
  {Atat\"ure}},\ }\href {\doibase 10.1103/PhysRevLett.108.093602} {\bibfield
  {journal} {\bibinfo  {journal} {Phys. Rev. Lett.}\ }\textbf {\bibinfo
  {volume} {108}},\ \bibinfo {pages} {093602} (\bibinfo {year}
  {2012})}\BibitemShut {NoStop}%
\bibitem [{\citenamefont {{He}}\ \emph {et~al.}(2013)\citenamefont {{He}},
  \citenamefont {{He}}, \citenamefont {{Wei}}, \citenamefont {{Wu}},
  \citenamefont {{Atat{\"u}re}}, \citenamefont {{Schneider}}, \citenamefont
  {{H{\"o}fling}}, \citenamefont {{Kamp}}, \citenamefont {{Lu}},\ and\
  \citenamefont {{Pan}}}]{He2013}%
  \BibitemOpen
  \bibfield  {author} {\bibinfo {author} {\bibfnamefont {Y.-M.}\ \bibnamefont
  {{He}}}, \bibinfo {author} {\bibfnamefont {Y.}~\bibnamefont {{He}}}, \bibinfo
  {author} {\bibfnamefont {Y.-J.}\ \bibnamefont {{Wei}}}, \bibinfo {author}
  {\bibfnamefont {D.}~\bibnamefont {{Wu}}}, \bibinfo {author} {\bibfnamefont
  {M.}~\bibnamefont {{Atat{\"u}re}}}, \bibinfo {author} {\bibfnamefont
  {C.}~\bibnamefont {{Schneider}}}, \bibinfo {author} {\bibfnamefont
  {S.}~\bibnamefont {{H{\"o}fling}}}, \bibinfo {author} {\bibfnamefont
  {M.}~\bibnamefont {{Kamp}}}, \bibinfo {author} {\bibfnamefont {C.-Y.}\
  \bibnamefont {{Lu}}}, \ and\ \bibinfo {author} {\bibfnamefont {J.-W.}\
  \bibnamefont {{Pan}}},\ }\href {\doibase 10.1038/nnano.2012.262} {\bibfield
  {journal} {\bibinfo  {journal} {Nature Nanotechnology}\ }\textbf {\bibinfo
  {volume} {8}},\ \bibinfo {pages} {213} (\bibinfo {year} {2013})}\BibitemShut
  {NoStop}%
\bibitem [{\citenamefont {Lodahl}\ \emph {et~al.}(2015)\citenamefont {Lodahl},
  \citenamefont {Mahmoodian},\ and\ \citenamefont {Stobbe}}]{Lodahl2015}%
  \BibitemOpen
  \bibfield  {author} {\bibinfo {author} {\bibfnamefont {P.}~\bibnamefont
  {Lodahl}}, \bibinfo {author} {\bibfnamefont {S.}~\bibnamefont {Mahmoodian}},
  \ and\ \bibinfo {author} {\bibfnamefont {S.}~\bibnamefont {Stobbe}},\ }\href
  {\doibase 10.1103/RevModPhys.87.347} {\bibfield  {journal} {\bibinfo
  {journal} {Rev. Mod. Phys.}\ }\textbf {\bibinfo {volume} {87}},\ \bibinfo
  {pages} {347} (\bibinfo {year} {2015})}\BibitemShut {NoStop}%
\bibitem [{\citenamefont {Ding}\ \emph {et~al.}(2016)\citenamefont {Ding},
  \citenamefont {He}, \citenamefont {Duan}, \citenamefont {Gregersen},
  \citenamefont {Chen}, \citenamefont {Unsleber}, \citenamefont {Maier},
  \citenamefont {Schneider}, \citenamefont {Kamp}, \citenamefont {H\"ofling},
  \citenamefont {Lu},\ and\ \citenamefont {Pan}}]{Ding2016}%
  \BibitemOpen
  \bibfield  {author} {\bibinfo {author} {\bibfnamefont {X.}~\bibnamefont
  {Ding}}, \bibinfo {author} {\bibfnamefont {Y.}~\bibnamefont {He}}, \bibinfo
  {author} {\bibfnamefont {Z.-C.}\ \bibnamefont {Duan}}, \bibinfo {author}
  {\bibfnamefont {N.}~\bibnamefont {Gregersen}}, \bibinfo {author}
  {\bibfnamefont {M.-C.}\ \bibnamefont {Chen}}, \bibinfo {author}
  {\bibfnamefont {S.}~\bibnamefont {Unsleber}}, \bibinfo {author}
  {\bibfnamefont {S.}~\bibnamefont {Maier}}, \bibinfo {author} {\bibfnamefont
  {C.}~\bibnamefont {Schneider}}, \bibinfo {author} {\bibfnamefont
  {M.}~\bibnamefont {Kamp}}, \bibinfo {author} {\bibfnamefont {S.}~\bibnamefont
  {H\"ofling}}, \bibinfo {author} {\bibfnamefont {C.-Y.}\ \bibnamefont {Lu}}, \
  and\ \bibinfo {author} {\bibfnamefont {J.-W.}\ \bibnamefont {Pan}},\ }\href
  {\doibase 10.1103/PhysRevLett.116.020401} {\bibfield  {journal} {\bibinfo
  {journal} {Phys. Rev. Lett.}\ }\textbf {\bibinfo {volume} {116}},\ \bibinfo
  {pages} {020401} (\bibinfo {year} {2016})}\BibitemShut {NoStop}%
\bibitem [{\citenamefont {Kim}(2008)}]{Kim2008}%
  \BibitemOpen
  \bibfield  {author} {\bibinfo {author} {\bibfnamefont {M.~S.}\ \bibnamefont
  {Kim}},\ }\href {http://stacks.iop.org/0953-4075/41/i=13/a=133001} {\bibfield
   {journal} {\bibinfo  {journal} {Journal of Physics B: Atomic, Molecular and
  Optical Physics}\ }\textbf {\bibinfo {volume} {41}},\ \bibinfo {pages}
  {133001} (\bibinfo {year} {2008})}\BibitemShut {NoStop}%
\bibitem [{\citenamefont {Yukawa}\ \emph {et~al.}(2013)\citenamefont {Yukawa},
  \citenamefont {Miyata}, \citenamefont {Yonezawa}, \citenamefont {Marek},
  \citenamefont {Filip},\ and\ \citenamefont {Furusawa}}]{Yukawa2013}%
  \BibitemOpen
  \bibfield  {author} {\bibinfo {author} {\bibfnamefont {M.}~\bibnamefont
  {Yukawa}}, \bibinfo {author} {\bibfnamefont {K.}~\bibnamefont {Miyata}},
  \bibinfo {author} {\bibfnamefont {H.}~\bibnamefont {Yonezawa}}, \bibinfo
  {author} {\bibfnamefont {P.}~\bibnamefont {Marek}}, \bibinfo {author}
  {\bibfnamefont {R.}~\bibnamefont {Filip}}, \ and\ \bibinfo {author}
  {\bibfnamefont {A.}~\bibnamefont {Furusawa}},\ }\href {\doibase
  10.1103/PhysRevA.88.053816} {\bibfield  {journal} {\bibinfo  {journal} {Phys.
  Rev. A}\ }\textbf {\bibinfo {volume} {88}},\ \bibinfo {pages} {053816}
  (\bibinfo {year} {2013})}\BibitemShut {NoStop}%
\bibitem [{\citenamefont {Marshall}\ \emph {et~al.}(2015)\citenamefont
  {Marshall}, \citenamefont {Pooser}, \citenamefont {Siopsis},\ and\
  \citenamefont {Weedbrook}}]{Marshall2015}%
  \BibitemOpen
  \bibfield  {author} {\bibinfo {author} {\bibfnamefont {K.}~\bibnamefont
  {Marshall}}, \bibinfo {author} {\bibfnamefont {R.}~\bibnamefont {Pooser}},
  \bibinfo {author} {\bibfnamefont {G.}~\bibnamefont {Siopsis}}, \ and\
  \bibinfo {author} {\bibfnamefont {C.}~\bibnamefont {Weedbrook}},\ }\href
  {\doibase 10.1103/PhysRevA.91.032321} {\bibfield  {journal} {\bibinfo
  {journal} {Phys. Rev. A}\ }\textbf {\bibinfo {volume} {91}},\ \bibinfo
  {pages} {032321} (\bibinfo {year} {2015})}\BibitemShut {NoStop}%
\bibitem [{\citenamefont {Arzani}\ \emph {et~al.}(2017)\citenamefont {Arzani},
  \citenamefont {Treps},\ and\ \citenamefont {Ferrini}}]{Arzani2017}%
  \BibitemOpen
  \bibfield  {author} {\bibinfo {author} {\bibfnamefont {F.}~\bibnamefont
  {Arzani}}, \bibinfo {author} {\bibfnamefont {N.}~\bibnamefont {Treps}}, \
  and\ \bibinfo {author} {\bibfnamefont {G.}~\bibnamefont {Ferrini}},\ }\href
  {\doibase 10.1103/PhysRevA.95.052352} {\bibfield  {journal} {\bibinfo
  {journal} {Phys. Rev. A}\ }\textbf {\bibinfo {volume} {95}},\ \bibinfo
  {pages} {052352} (\bibinfo {year} {2017})}\BibitemShut {NoStop}%
\bibitem [{\citenamefont {Wiseman}\ and\ \citenamefont
  {Milburn}(2010)}]{Wiseman}%
  \BibitemOpen
  \bibfield  {author} {\bibinfo {author} {\bibfnamefont {H.}~\bibnamefont
  {Wiseman}}\ and\ \bibinfo {author} {\bibfnamefont {G.}~\bibnamefont
  {Milburn}},\ }\href@noop {} {\emph {\bibinfo {title} {Quantum Measurement and
  Control}}}\ (\bibinfo  {publisher} {Cambridge University Press},\ \bibinfo
  {year} {2010})\BibitemShut {NoStop}%
\bibitem [{\citenamefont {Brun}(2002)}]{Brun2002}%
  \BibitemOpen
  \bibfield  {author} {\bibinfo {author} {\bibfnamefont {T.~A.}\ \bibnamefont
  {Brun}},\ }\href {\doibase 10.1119/1.1475328} {\bibfield  {journal} {\bibinfo
   {journal} {American Journal of Physics}\ }\textbf {\bibinfo {volume} {70}},\
  \bibinfo {pages} {719} (\bibinfo {year} {2002})}\BibitemShut {NoStop}%
\bibitem [{\citenamefont {Lvovsky}(2004)}]{Lvovsky2004}%
  \BibitemOpen
  \bibfield  {author} {\bibinfo {author} {\bibfnamefont {A.~I.}\ \bibnamefont
  {Lvovsky}},\ }\href {http://stacks.iop.org/1464-4266/6/i=6/a=014} {\bibfield
  {journal} {\bibinfo  {journal} {Journal of Optics B: Quantum and
  Semiclassical Optics}\ }\textbf {\bibinfo {volume} {6}},\ \bibinfo {pages}
  {S556} (\bibinfo {year} {2004})}\BibitemShut {NoStop}%
\bibitem [{\citenamefont {Lvovsky}\ and\ \citenamefont
  {Raymer}(2009)}]{Lvovsky2009}%
  \BibitemOpen
  \bibfield  {author} {\bibinfo {author} {\bibfnamefont {A.~I.}\ \bibnamefont
  {Lvovsky}}\ and\ \bibinfo {author} {\bibfnamefont {M.~G.}\ \bibnamefont
  {Raymer}},\ }\href {\doibase 10.1103/RevModPhys.81.299} {\bibfield  {journal}
  {\bibinfo  {journal} {Rev. Mod. Phys.}\ }\textbf {\bibinfo {volume} {81}},\
  \bibinfo {pages} {299} (\bibinfo {year} {2009})}\BibitemShut {NoStop}%
\bibitem [{\citenamefont {Pichler}\ and\ \citenamefont
  {Zoller}(2016)}]{Pichler2016}%
  \BibitemOpen
  \bibfield  {author} {\bibinfo {author} {\bibfnamefont {H.}~\bibnamefont
  {Pichler}}\ and\ \bibinfo {author} {\bibfnamefont {P.}~\bibnamefont
  {Zoller}},\ }\href {\doibase 10.1103/PhysRevLett.116.093601} {\bibfield
  {journal} {\bibinfo  {journal} {Phys. Rev. Lett.}\ }\textbf {\bibinfo
  {volume} {116}},\ \bibinfo {pages} {093601} (\bibinfo {year}
  {2016})}\BibitemShut {NoStop}%
\bibitem [{\citenamefont {Gardiner}\ and\ \citenamefont
  {Collett}(1985)}]{Gardiner1985}%
  \BibitemOpen
  \bibfield  {author} {\bibinfo {author} {\bibfnamefont {C.~W.}\ \bibnamefont
  {Gardiner}}\ and\ \bibinfo {author} {\bibfnamefont {M.~J.}\ \bibnamefont
  {Collett}},\ }\href {\doibase 10.1103/PhysRevA.31.3761} {\bibfield  {journal}
  {\bibinfo  {journal} {Phys. Rev. A}\ }\textbf {\bibinfo {volume} {31}},\
  \bibinfo {pages} {3761} (\bibinfo {year} {1985})}\BibitemShut {NoStop}%
\bibitem [{EPA()}]{EPAPS}%
  \BibitemOpen
  \href@noop {} {}\bibinfo {howpublished} {See supplementary material in this
  submission.}\BibitemShut {Stop}%
\bibitem [{\citenamefont {Hudson}(1974)}]{Hudson1974}%
  \BibitemOpen
  \bibfield  {author} {\bibinfo {author} {\bibfnamefont {R.}~\bibnamefont
  {Hudson}},\ }\href@noop {} {\bibfield  {journal} {\bibinfo  {journal}
  {Reports on Mathematical Physics}\ }\textbf {\bibinfo {volume} {6}},\
  \bibinfo {pages} {249 } (\bibinfo {year} {1974})}\BibitemShut {NoStop}%
\bibitem [{\citenamefont {Loudon}(1983)}]{Loudon1983}%
  \BibitemOpen
  \bibfield  {author} {\bibinfo {author} {\bibfnamefont {R.}~\bibnamefont
  {Loudon}},\ }\href {https://books.google.se/books?id=\_FTwAAAAMAAJ} {\emph
  {\bibinfo {title} {The quantum theory of light}}},\ Oxford science
  publications\ (\bibinfo  {publisher} {Clarendon Press},\ \bibinfo {year}
  {1983})\BibitemShut {NoStop}%
\bibitem [{\citenamefont {Strandberg}(2017)}]{Strandberg2017}%
  \BibitemOpen
  \bibfield  {author} {\bibinfo {author} {\bibfnamefont {I.}~\bibnamefont
  {Strandberg}},\ }\emph {\bibinfo {title} {Quantum state tomography of 1D
  resonance fluorescence}},\ \href
  {http://publications.lib.chalmers.se/records/fulltext/252882/252882.pdf}
  {Master's thesis},\ \bibinfo  {school} {Chalmers University of Technology}
  (\bibinfo {year} {2017})\BibitemShut {NoStop}%
\bibitem [{\citenamefont {{Moyal}}\ and\ \citenamefont
  {{Bartlett}}(1949)}]{Moyal1949}%
  \BibitemOpen
  \bibfield  {author} {\bibinfo {author} {\bibfnamefont {J.~E.}\ \bibnamefont
  {{Moyal}}}\ and\ \bibinfo {author} {\bibfnamefont {M.~S.}\ \bibnamefont
  {{Bartlett}}},\ }\href {\doibase 10.1017/S0305004100000487} {\bibfield
  {journal} {\bibinfo  {journal} {Proceedings of the Cambridge Philosophical
  Society}\ }\textbf {\bibinfo {volume} {45}},\ \bibinfo {pages} {99} (\bibinfo
  {year} {1949})}\BibitemShut {NoStop}%
\bibitem [{\citenamefont {{Lindkvist}}\ and\ \citenamefont
  {{Johansson}}(2014)}]{Joel2014}%
  \BibitemOpen
  \bibfield  {author} {\bibinfo {author} {\bibfnamefont {J.}~\bibnamefont
  {{Lindkvist}}}\ and\ \bibinfo {author} {\bibfnamefont {G.}~\bibnamefont
  {{Johansson}}},\ }\href {\doibase 10.1088/1367-2630/16/5/055018} {\bibfield
  {journal} {\bibinfo  {journal} {New Journal of Physics}\ }\textbf {\bibinfo
  {volume} {16}},\ \bibinfo {eid} {055018} (\bibinfo {year}
  {2014})}\BibitemShut {NoStop}%
\bibitem [{\citenamefont {Kenfack}\ and\ \citenamefont
  {\.Zyczkowski}(2004)}]{Kenfack2004}%
  \BibitemOpen
  \bibfield  {author} {\bibinfo {author} {\bibfnamefont {A.}~\bibnamefont
  {Kenfack}}\ and\ \bibinfo {author} {\bibfnamefont {K.}~\bibnamefont
  {\.Zyczkowski}},\ }\href {http://stacks.iop.org/1464-4266/6/i=10/a=003}
  {\bibfield  {journal} {\bibinfo  {journal} {Journal of Optics B: Quantum and
  Semiclassical Optics}\ }\textbf {\bibinfo {volume} {6}},\ \bibinfo {pages}
  {396} (\bibinfo {year} {2004})}\BibitemShut {NoStop}%
\bibitem [{\citenamefont {Albarelli}\ \emph {et~al.}(2016)\citenamefont
  {Albarelli}, \citenamefont {Ferraro}, \citenamefont {Paternostro},\ and\
  \citenamefont {Paris}}]{Albarelli2016}%
  \BibitemOpen
  \bibfield  {author} {\bibinfo {author} {\bibfnamefont {F.}~\bibnamefont
  {Albarelli}}, \bibinfo {author} {\bibfnamefont {A.}~\bibnamefont {Ferraro}},
  \bibinfo {author} {\bibfnamefont {M.}~\bibnamefont {Paternostro}}, \ and\
  \bibinfo {author} {\bibfnamefont {M.~G.~A.}\ \bibnamefont {Paris}},\
  }\href@noop {} {\bibfield  {journal} {\bibinfo  {journal} {Phys. Rev. A}\
  }\textbf {\bibinfo {volume} {93}},\ \bibinfo {pages} {032112} (\bibinfo
  {year} {2016})}\BibitemShut {NoStop}%
\bibitem [{\citenamefont {Joana}\ \emph {et~al.}(2016)\citenamefont {Joana},
  \citenamefont {van Loock}, \citenamefont {Deng},\ and\ \citenamefont
  {Byrnes}}]{Joana2016}%
  \BibitemOpen
  \bibfield  {author} {\bibinfo {author} {\bibfnamefont {C.}~\bibnamefont
  {Joana}}, \bibinfo {author} {\bibfnamefont {P.}~\bibnamefont {van Loock}},
  \bibinfo {author} {\bibfnamefont {H.}~\bibnamefont {Deng}}, \ and\ \bibinfo
  {author} {\bibfnamefont {T.}~\bibnamefont {Byrnes}},\ }\href {\doibase
  10.1103/PhysRevA.94.063802} {\bibfield  {journal} {\bibinfo  {journal} {Phys.
  Rev. A}\ }\textbf {\bibinfo {volume} {94}},\ \bibinfo {pages} {063802}
  (\bibinfo {year} {2016})}\BibitemShut {NoStop}%
\bibitem [{\citenamefont {Mallet}\ \emph {et~al.}(2011)\citenamefont {Mallet},
  \citenamefont {Castellanos-Beltran}, \citenamefont {Ku}, \citenamefont
  {Glancy}, \citenamefont {Knill}, \citenamefont {Irwin}, \citenamefont
  {Hilton}, \citenamefont {Vale},\ and\ \citenamefont {Lehnert}}]{Mallet2011}%
  \BibitemOpen
  \bibfield  {author} {\bibinfo {author} {\bibfnamefont {F.}~\bibnamefont
  {Mallet}}, \bibinfo {author} {\bibfnamefont {M.~A.}\ \bibnamefont
  {Castellanos-Beltran}}, \bibinfo {author} {\bibfnamefont {H.~S.}\
  \bibnamefont {Ku}}, \bibinfo {author} {\bibfnamefont {S.}~\bibnamefont
  {Glancy}}, \bibinfo {author} {\bibfnamefont {E.}~\bibnamefont {Knill}},
  \bibinfo {author} {\bibfnamefont {K.~D.}\ \bibnamefont {Irwin}}, \bibinfo
  {author} {\bibfnamefont {G.~C.}\ \bibnamefont {Hilton}}, \bibinfo {author}
  {\bibfnamefont {L.~R.}\ \bibnamefont {Vale}}, \ and\ \bibinfo {author}
  {\bibfnamefont {K.~W.}\ \bibnamefont {Lehnert}},\ }\href {\doibase
  10.1103/PhysRevLett.106.220502} {\bibfield  {journal} {\bibinfo  {journal}
  {Phys. Rev. Lett.}\ }\textbf {\bibinfo {volume} {106}},\ \bibinfo {pages}
  {220502} (\bibinfo {year} {2011})}\BibitemShut {NoStop}%
\bibitem [{\citenamefont {Menzel}\ \emph {et~al.}(2010)\citenamefont {Menzel},
  \citenamefont {Deppe}, \citenamefont {Mariantoni}, \citenamefont
  {Araque~Caballero}, \citenamefont {Baust}, \citenamefont {Niemczyk},
  \citenamefont {Hoffmann}, \citenamefont {Marx}, \citenamefont {Solano},\ and\
  \citenamefont {Gross}}]{Menzel2010}%
  \BibitemOpen
  \bibfield  {author} {\bibinfo {author} {\bibfnamefont {E.~P.}\ \bibnamefont
  {Menzel}}, \bibinfo {author} {\bibfnamefont {F.}~\bibnamefont {Deppe}},
  \bibinfo {author} {\bibfnamefont {M.}~\bibnamefont {Mariantoni}}, \bibinfo
  {author} {\bibfnamefont {M.~A.}\ \bibnamefont {Araque~Caballero}}, \bibinfo
  {author} {\bibfnamefont {A.}~\bibnamefont {Baust}}, \bibinfo {author}
  {\bibfnamefont {T.}~\bibnamefont {Niemczyk}}, \bibinfo {author}
  {\bibfnamefont {E.}~\bibnamefont {Hoffmann}}, \bibinfo {author}
  {\bibfnamefont {A.}~\bibnamefont {Marx}}, \bibinfo {author} {\bibfnamefont
  {E.}~\bibnamefont {Solano}}, \ and\ \bibinfo {author} {\bibfnamefont
  {R.}~\bibnamefont {Gross}},\ }\href {\doibase 10.1103/PhysRevLett.105.100401}
  {\bibfield  {journal} {\bibinfo  {journal} {Phys. Rev. Lett.}\ }\textbf
  {\bibinfo {volume} {105}},\ \bibinfo {pages} {100401} (\bibinfo {year}
  {2010})}\BibitemShut {NoStop}%
\bibitem [{\citenamefont {da~Silva}\ \emph {et~al.}(2010)\citenamefont
  {da~Silva}, \citenamefont {Bozyigit}, \citenamefont {Wallraff},\ and\
  \citenamefont {Blais}}]{daSilva2010}%
  \BibitemOpen
  \bibfield  {author} {\bibinfo {author} {\bibfnamefont {M.~P.}\ \bibnamefont
  {da~Silva}}, \bibinfo {author} {\bibfnamefont {D.}~\bibnamefont {Bozyigit}},
  \bibinfo {author} {\bibfnamefont {A.}~\bibnamefont {Wallraff}}, \ and\
  \bibinfo {author} {\bibfnamefont {A.}~\bibnamefont {Blais}},\ }\href
  {\doibase 10.1103/PhysRevA.82.043804} {\bibfield  {journal} {\bibinfo
  {journal} {Phys. Rev. A}\ }\textbf {\bibinfo {volume} {82}},\ \bibinfo
  {pages} {043804} (\bibinfo {year} {2010})}\BibitemShut {NoStop}%
\bibitem [{\citenamefont {Menzel}\ \emph {et~al.}(2012)\citenamefont {Menzel},
  \citenamefont {Di~Candia}, \citenamefont {Deppe}, \citenamefont {Eder},
  \citenamefont {Zhong}, \citenamefont {Ihmig}, \citenamefont {Haeberlein},
  \citenamefont {Baust}, \citenamefont {Hoffmann}, \citenamefont {Ballester},
  \citenamefont {Inomata}, \citenamefont {Yamamoto}, \citenamefont {Nakamura},
  \citenamefont {Solano}, \citenamefont {Marx},\ and\ \citenamefont
  {Gross}}]{Menzel2012}%
  \BibitemOpen
  \bibfield  {author} {\bibinfo {author} {\bibfnamefont {E.~P.}\ \bibnamefont
  {Menzel}}, \bibinfo {author} {\bibfnamefont {R.}~\bibnamefont {Di~Candia}},
  \bibinfo {author} {\bibfnamefont {F.}~\bibnamefont {Deppe}}, \bibinfo
  {author} {\bibfnamefont {P.}~\bibnamefont {Eder}}, \bibinfo {author}
  {\bibfnamefont {L.}~\bibnamefont {Zhong}}, \bibinfo {author} {\bibfnamefont
  {M.}~\bibnamefont {Ihmig}}, \bibinfo {author} {\bibfnamefont
  {M.}~\bibnamefont {Haeberlein}}, \bibinfo {author} {\bibfnamefont
  {A.}~\bibnamefont {Baust}}, \bibinfo {author} {\bibfnamefont
  {E.}~\bibnamefont {Hoffmann}}, \bibinfo {author} {\bibfnamefont
  {D.}~\bibnamefont {Ballester}}, \bibinfo {author} {\bibfnamefont
  {K.}~\bibnamefont {Inomata}}, \bibinfo {author} {\bibfnamefont
  {T.}~\bibnamefont {Yamamoto}}, \bibinfo {author} {\bibfnamefont
  {Y.}~\bibnamefont {Nakamura}}, \bibinfo {author} {\bibfnamefont
  {E.}~\bibnamefont {Solano}}, \bibinfo {author} {\bibfnamefont
  {A.}~\bibnamefont {Marx}}, \ and\ \bibinfo {author} {\bibfnamefont
  {R.}~\bibnamefont {Gross}},\ }\href {\doibase 10.1103/PhysRevLett.109.250502}
  {\bibfield  {journal} {\bibinfo  {journal} {Phys. Rev. Lett.}\ }\textbf
  {\bibinfo {volume} {109}},\ \bibinfo {pages} {250502} (\bibinfo {year}
  {2012})}\BibitemShut {NoStop}%
\bibitem [{\citenamefont {Fedorov}\ \emph {et~al.}(2016)\citenamefont
  {Fedorov}, \citenamefont {Zhong}, \citenamefont {Pogorzalek}, \citenamefont
  {Eder}, \citenamefont {Fischer}, \citenamefont {Goetz}, \citenamefont {Xie},
  \citenamefont {Wulschner}, \citenamefont {Inomata}, \citenamefont {Yamamoto},
  \citenamefont {Nakamura}, \citenamefont {Di~Candia}, \citenamefont
  {Las~Heras}, \citenamefont {Sanz}, \citenamefont {Solano}, \citenamefont
  {Menzel}, \citenamefont {Deppe}, \citenamefont {Marx},\ and\ \citenamefont
  {Gross}}]{Fedorov2016}%
  \BibitemOpen
  \bibfield  {author} {\bibinfo {author} {\bibfnamefont {K.~G.}\ \bibnamefont
  {Fedorov}}, \bibinfo {author} {\bibfnamefont {L.}~\bibnamefont {Zhong}},
  \bibinfo {author} {\bibfnamefont {S.}~\bibnamefont {Pogorzalek}}, \bibinfo
  {author} {\bibfnamefont {P.}~\bibnamefont {Eder}}, \bibinfo {author}
  {\bibfnamefont {M.}~\bibnamefont {Fischer}}, \bibinfo {author} {\bibfnamefont
  {J.}~\bibnamefont {Goetz}}, \bibinfo {author} {\bibfnamefont
  {E.}~\bibnamefont {Xie}}, \bibinfo {author} {\bibfnamefont {F.}~\bibnamefont
  {Wulschner}}, \bibinfo {author} {\bibfnamefont {K.}~\bibnamefont {Inomata}},
  \bibinfo {author} {\bibfnamefont {T.}~\bibnamefont {Yamamoto}}, \bibinfo
  {author} {\bibfnamefont {Y.}~\bibnamefont {Nakamura}}, \bibinfo {author}
  {\bibfnamefont {R.}~\bibnamefont {Di~Candia}}, \bibinfo {author}
  {\bibfnamefont {U.}~\bibnamefont {Las~Heras}}, \bibinfo {author}
  {\bibfnamefont {M.}~\bibnamefont {Sanz}}, \bibinfo {author} {\bibfnamefont
  {E.}~\bibnamefont {Solano}}, \bibinfo {author} {\bibfnamefont {E.~P.}\
  \bibnamefont {Menzel}}, \bibinfo {author} {\bibfnamefont {F.}~\bibnamefont
  {Deppe}}, \bibinfo {author} {\bibfnamefont {A.}~\bibnamefont {Marx}}, \ and\
  \bibinfo {author} {\bibfnamefont {R.}~\bibnamefont {Gross}},\ }\href
  {\doibase 10.1103/PhysRevLett.117.020502} {\bibfield  {journal} {\bibinfo
  {journal} {Phys. Rev. Lett.}\ }\textbf {\bibinfo {volume} {117}},\ \bibinfo
  {pages} {020502} (\bibinfo {year} {2016})}\BibitemShut {NoStop}%
\bibitem [{\citenamefont {Boehm}\ \emph {et~al.}(2017)\citenamefont {Boehm},
  \citenamefont {Grosse}, \citenamefont {Kolarczik}, \citenamefont {Herzog},
  \citenamefont {Owschimikow},\ and\ \citenamefont {Woggon}}]{Boehm2017}%
  \BibitemOpen
  \bibfield  {author} {\bibinfo {author} {\bibfnamefont {F.}~\bibnamefont
  {Boehm}}, \bibinfo {author} {\bibfnamefont {N.~B.}\ \bibnamefont {Grosse}},
  \bibinfo {author} {\bibfnamefont {M.}~\bibnamefont {Kolarczik}}, \bibinfo
  {author} {\bibfnamefont {B.}~\bibnamefont {Herzog}}, \bibinfo {author}
  {\bibfnamefont {N.}~\bibnamefont {Owschimikow}}, \ and\ \bibinfo {author}
  {\bibfnamefont {U.~K.}\ \bibnamefont {Woggon}},\ }in\ \href {\doibase
  10.1364/NLO.2017.NTh2A.1} {\emph {\bibinfo {booktitle} {Nonlinear Optics}}}\
  (\bibinfo  {publisher} {Optical Society of America},\ \bibinfo {year}
  {2017})\ p.\ \bibinfo {pages} {NTh2A.1}\BibitemShut {NoStop}%
\bibitem [{\citenamefont {{Albarelli}}\ \emph {et~al.}(2018)\citenamefont
  {{Albarelli}}, \citenamefont {{Genoni}}, \citenamefont {{Paris}},\ and\
  \citenamefont {{Ferraro}}}]{Albarelli2018}%
  \BibitemOpen
  \bibfield  {author} {\bibinfo {author} {\bibfnamefont {F.}~\bibnamefont
  {{Albarelli}}}, \bibinfo {author} {\bibfnamefont {M.~G.}\ \bibnamefont
  {{Genoni}}}, \bibinfo {author} {\bibfnamefont {M.~G.~A.}\ \bibnamefont
  {{Paris}}}, \ and\ \bibinfo {author} {\bibfnamefont {A.}~\bibnamefont
  {{Ferraro}}},\ }\href@noop {} {\bibfield  {journal} {\bibinfo  {journal}
  {ArXiv e-prints}\ } (\bibinfo {year} {2018})},\ \Eprint
  {http://arxiv.org/abs/1804.05763} {arXiv:1804.05763 [quant-ph]} \BibitemShut
  {NoStop}%
\end{thebibliography}%

\newpage
\begin{appendix}

\begin{widetext}

\section{Two-time correlation functions in the steady-state}
\label{app:twotime}

The steady-state (ss) two-time correlation $\langle \hat a_{\rm out}^\dagger (t) \hat a_{\rm out}(0) \rangle_{\rm ss}$ (Eq. (4) in the main text), can be related to correlations
of the TLS lowering $\hat \sigma_-$ and raising $\hat \sigma_+$
operators by the input-output
relation (Eq. (3) in the main text). In this way, we have
\begin{equation}\label{app:corr}
\langle \hat a_{\rm out}^\dagger (t) \hat a_{\rm out}(0) \rangle_{\rm ss} =
\Omega^2 + \Omega \sqrt{\gamma} \left( \langle \hat \sigma_+ \rangle_{\rm ss} + \rm{c.c.} \right)
+ \gamma \langle \hat \sigma_+ (t) \sigma_-(0) \rangle_{\rm ss} .
\end{equation}
The first step is to solve for the
vector $\vec \sigma = (\langle \hat \sigma_+ \rangle, \langle
\hat \sigma_- \rangle, \langle \hat \sigma_z \rangle )$  in the
steady-state. The solutions are
\begin{align}
\langle \hat \sigma_+ \rangle_{\rm ss} &= \langle \hat \sigma_- \rangle_{\rm ss} =
- \frac{2 \Omega/ \sqrt{\gamma}}{ 1 + 8 \Omega^2/ \gamma}, \label{app:bloch}\\
\langle \hat \sigma_z \rangle_{\rm ss} &= -\frac{1}{1 + 8 \Omega^2/ \gamma} .
\end{align}
From here, the two-time correlation $\langle \hat \sigma_+ (t) \hat
\sigma_-(0)  \rangle_{\rm ss}$ of the TLS operators
can be calculated by
using the quantum regression theorem~\cite{Walls}. The result is
\begin{equation}\label{app:corr2}
 \langle \hat \sigma_+ (t) \hat \sigma_-(0)  \rangle_{\rm ss} = \frac{2 \Omega^2/ \gamma}{1 + 8 \Omega^2/ \gamma} \exp \left( - \gamma t/2 \right)
+ \frac{\lambda_+}{\gamma}  \exp \left[ - \frac{\gamma t}{4} \left( 3 + \imi \sqrt{64  \Omega^2/\gamma -1}  \right) \right] + {\rm c.c.},
\end{equation}
with
\begin{align}
\lambda_+ &= \frac{\Omega^2  \left(-1 + \sqrt{1 -64\, \Omega^2 /\gamma  } \right) \left(16 \,\Omega^2 /\gamma - 1 + \sqrt{1 -64\, \Omega^2/\gamma } \right)}{2 ( 1+ 8 \Omega^2/\gamma)^2 \sqrt{1 -64\, \Omega^2 /\gamma} } .
\end{align}
Inserting~\eqref{app:bloch} and~\eqref{app:corr2} in~\eqref{app:corr} yields Eq. (3) in the main text.

\section{Wigner function of a displaced state}

The Wigner function of a state $\rho$ is defined as~\cite{Hillery1984}
\begin{equation}\label{app:Wigner1}
W(x,p) = \frac{1}{2\pi} \int_{-\infty}^\infty {\rm d}y \, {\rm e}^{\imi p y} \langle x + y/2 \vert \rho \vert x - y/2 \rangle .
\end{equation}
We now introduce the (unitary) displacement operator $\hat D(\alpha)$ on a bosonic
mode annihilated (created) by $\hat a$ ($\hat a^\dagger$) satisfying the
commutation relation $[\hat a, \hat a^\dagger] = 1$
\begin{equation}\label{app:Dop}
\hat D(\alpha) = \exp \left(\alpha \hat a^\dagger - \alpha^* \hat a \right).
\end{equation}
Its action on the annihilation operator is defined by
\begin{equation}
\hat D(\alpha)^\dagger \hat a  \hat D(\alpha) = \hat a + \alpha  .
\end{equation}
We can rewrite~\eqref{app:Dop} in terms of the quadrature operators
$\hat x = (\hat a^\dagger + \hat a)/2$ and $\hat p = \imi(\hat a^\dagger - \hat a)/2$:
\begin{equation}
\hat D(\alpha) = \exp \left[ \imi \left( \Im(\alpha)\, \hat x - \Re(\alpha)\, \hat p \right) \right].
\end{equation}
Its action on the position quadrature  eigenstates $\vert x \rangle$
($\hat x \vert x \rangle = x \vert x \rangle $) is
\begin{equation}\label{app:disp}
\hat D(\alpha) \vert x \rangle = \exp \left[ \imi\left( x + 3 \Re (\alpha)/2  \right) \right] \vert x + \Re (\alpha) \rangle .
\end{equation}
The displaced state $\rho_\alpha$ is defined by
\begin{equation}
\rho_\alpha = \hat D^\dagger (\alpha) \rho \hat D(\alpha) .
\end{equation}
Following the definition~\eqref{app:Wigner1}, the Wigner function of
the displaced state is
 \begin{align}
W_\alpha(x,p) =  \frac{1}{2\pi} \int_{-\infty}^\infty {\rm d}y \, {\rm e}^{ \imi p y} \langle x + y/2 \vert \hat D^\dagger(\alpha) \rho  \hat D(\alpha) \vert x - y/2 \rangle .
 \end{align}
 Using relation~\eqref{app:disp}, this reduces to
\begin{align}\label{app:Wigner2}
W_\alpha(x,p) =  \frac{1}{2\pi} \int_{-\infty}^\infty {\rm d}y \, {\rm e}^{\imi [p + \Im (\alpha)] y}  \langle x + \Re (\alpha) + y/2 \vert \rho \vert x + \Re (\alpha) - y/2 \rangle .
\end{align}
Comparing~\eqref{app:Wigner1} and~\eqref{app:Wigner2}, we see that
they are related by a translation, i.e., we go from~\eqref{app:Wigner2} to
\eqref{app:Wigner1} by shifting the  origin of phase space from $(0,0)$ to
$(\Re (\alpha), \Im (\alpha))$.
The  Wigner function is simply displaced by $\vert \alpha \vert^2$.

\section{Purity}

In Fig.~\ref{fig:apppurity}
we show the purity
of the filtered output field
as a function of the filter time $T$ and
for drive strengths $\Omega=0.2$, 0.35, 0.5 and 0.8 (in units of $\gamma =1$).
These correspond to the states studied in the main text [Cf. Fig. (4)].

\begin{figure}[t]
\begin{center}
\includegraphics[width=.6\columnwidth]{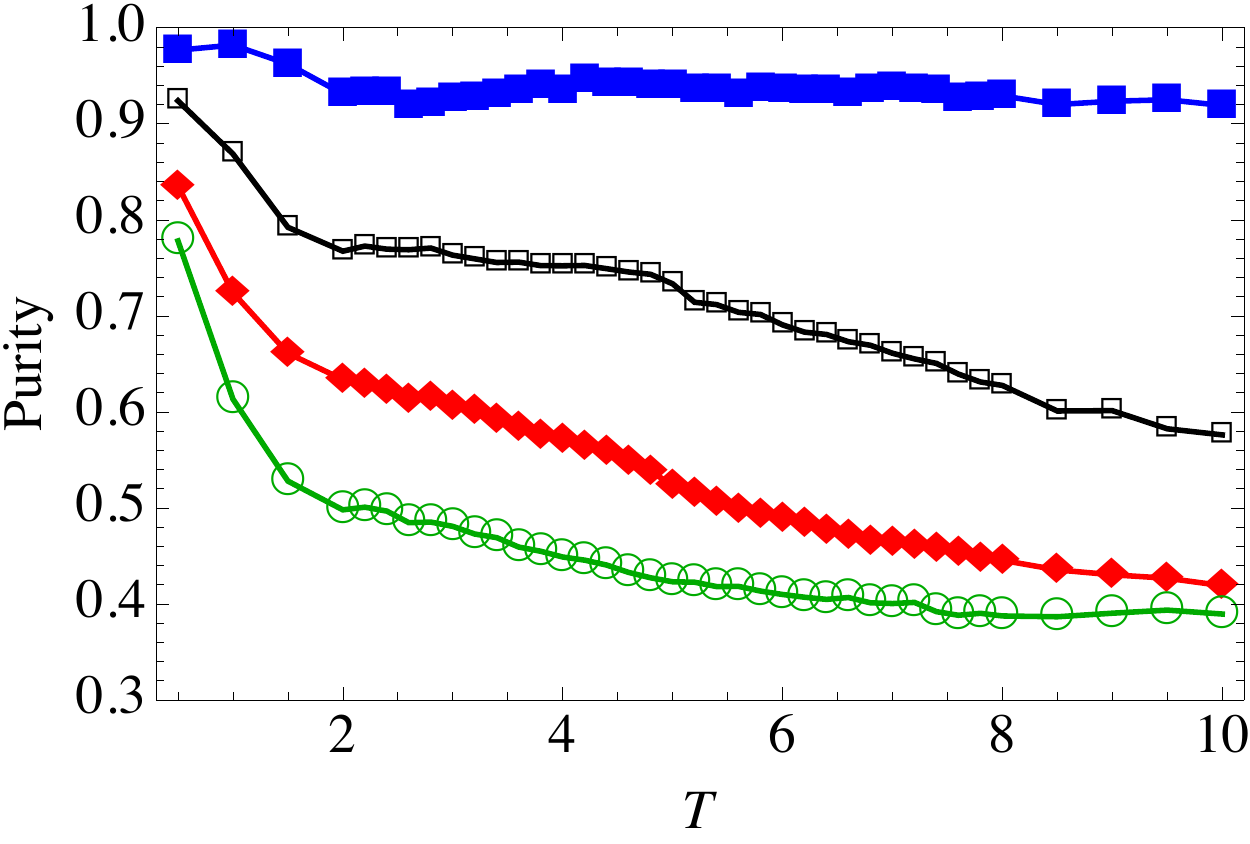}
\end{center}
\caption{(Color online)  Purity as a function of $T$ for
drive strengths $\Omega = 0.2$ (filled blue squares), $0.35$
(open black squares), $0.5$ (filled red diamonds)
and $0.8$ (open green circles).
Here we set $\gamma = 1$ as the unit, consequently $\Omega^* = 0.35$.}
\label{fig:apppurity}
\end{figure}

\section{Photon populations}

In Figs.~\ref{fig:apppop1},~\ref{fig:apppop2} and~\ref{fig:apppop3}
we show the photon populations
in the filtered output field
as a function of the filter time $T$ and
for drive strengths $\Omega=0.2$, 0.5 and 0.8 (in units of $\gamma =1$).
These populations correspond to the states studied in the main text [Cf. Fig. (4)].
\begin{figure}[t]
\begin{center}
\includegraphics[width=.6\columnwidth]{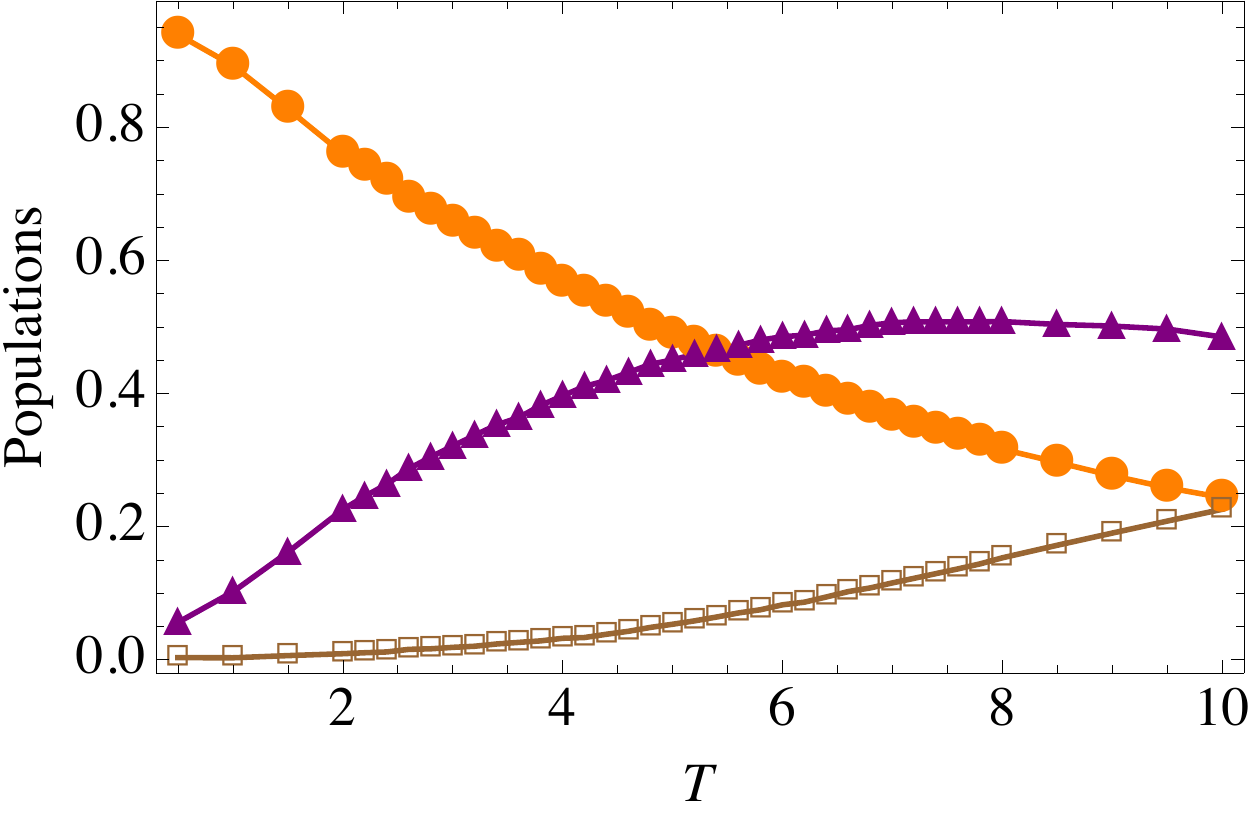}
\end{center}
\caption{(Color online)  Vacuum (filled orange circles), single- (filled purple triangles)
and two-photon
(open brown squares) populations
for the reconstructed state of the output field for a drive strength $\Omega = 0.2$ and as
a function of $T$. Here we set $\gamma =1$ as the unit.}
\label{fig:apppop1}
\end{figure}

\begin{figure}[t]
\begin{center}
\includegraphics[width=0.6\columnwidth]{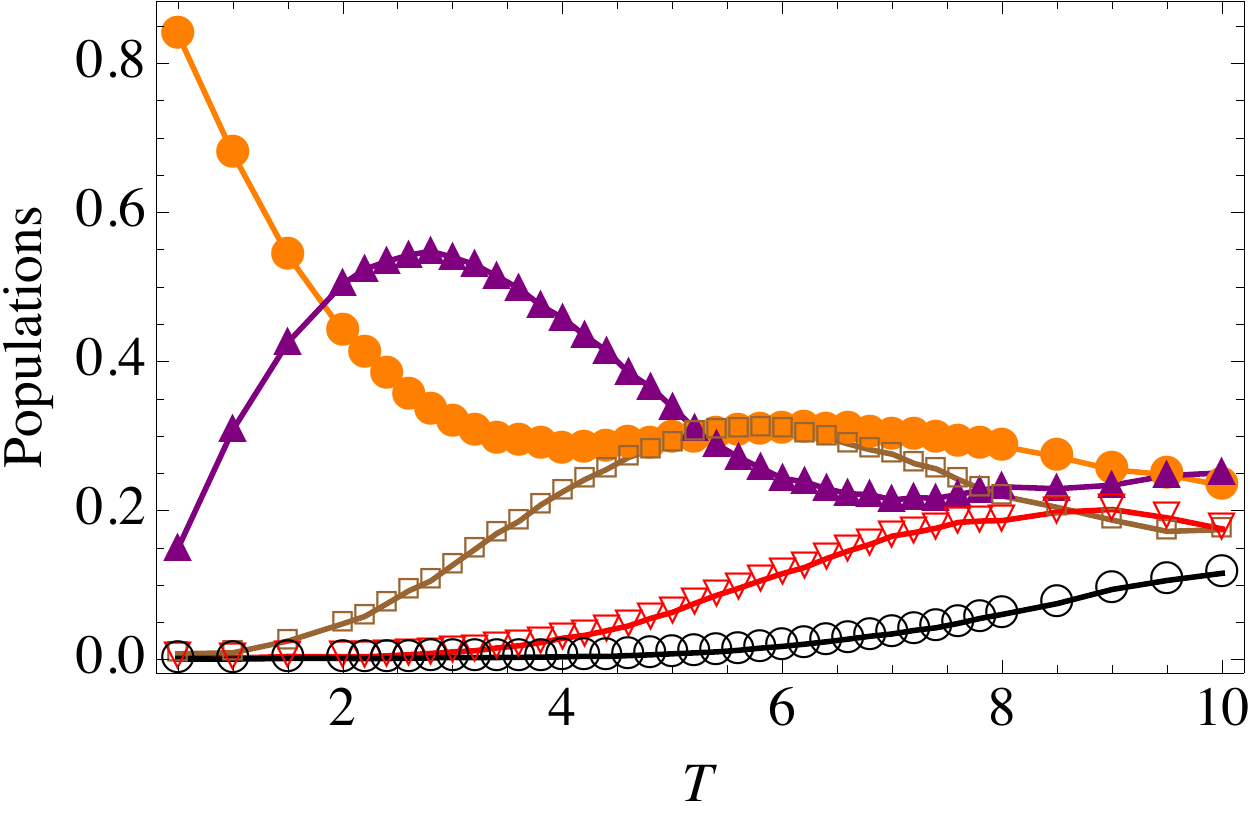}
\end{center}
\caption{(Color online)  Vacuum (filled orange circles), single- (filled purple triangles)
two- (open brown squares), three-  (open red inverted triangles) and
four-photon populations (open black circles)
for the reconstructed state of the output field for a drive strength $\Omega = 0.5$ and as
a function of $T$. Here we set $\gamma =1$ as the unit.}
\label{fig:apppop2}
\end{figure}

\begin{figure}[t]
\begin{center}
\includegraphics[width=.6\columnwidth]{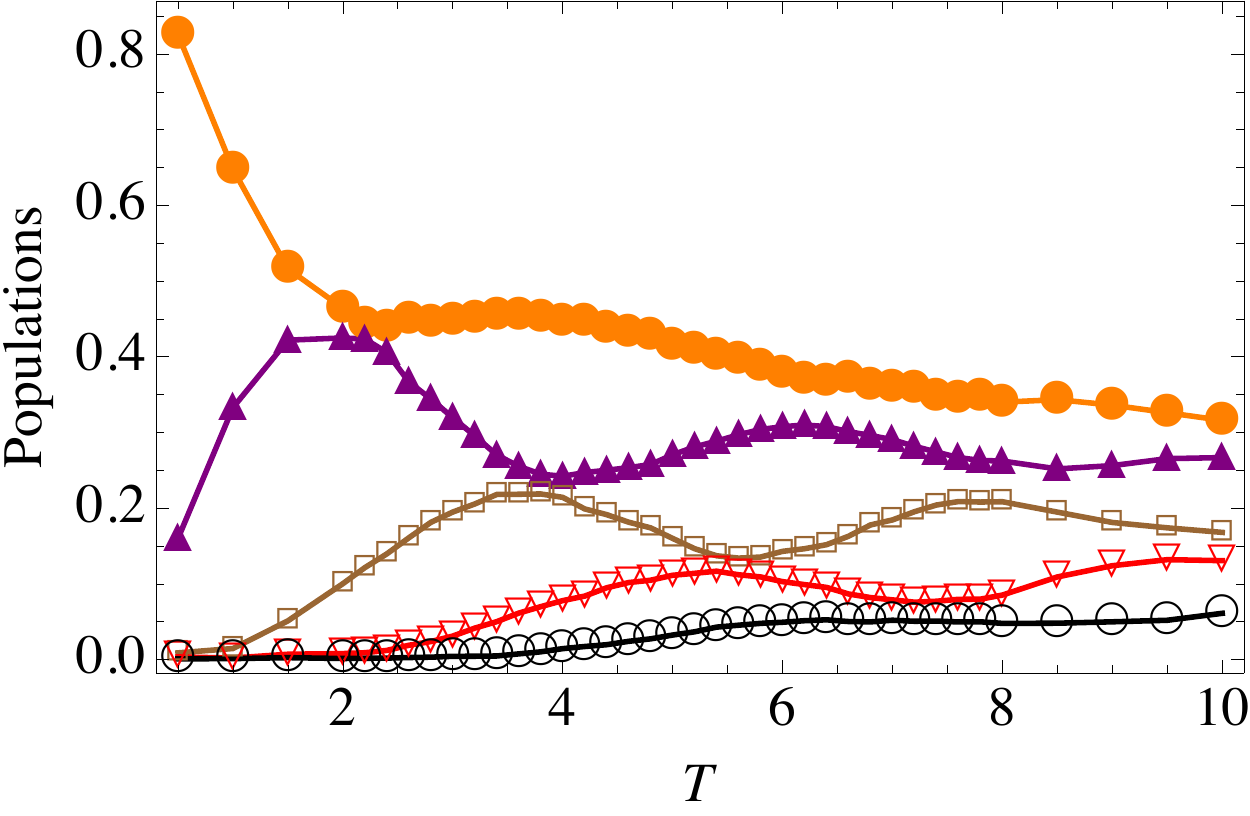}
\end{center}
\caption{(Color online)  Vacuum (filled orange circles), single- (filled purple triangles)
two- (open brown squares), three-  (open red inverted triangles) and
four-photon populations (open black circles)
for the reconstructed state of the output field for a drive strength $\Omega = 0.8$ and as
a function of $T$. Here we set $\gamma =1$ as the unit.}
\label{fig:apppop3}
\end{figure}

\section{Maximum Likelihood}

The goal of homodyne tomography is to infer the state $\rho$
of a single bosonic mode $\hat a$ with $[\hat a, \hat a^\dagger] = 1$.
In a homodyne detection experiment, the measured quantity is the
photocurrent, which is proportional to the expectation value
of the generalized quadrature observable
\begin{equation}
\hat X_\theta = \hat a^\dagger {\rm e}^{-i \theta} + \hat a {\rm e}^{+i \theta},
\end{equation}
with $\theta$ a real phase.
The quadrature operator is
a continuous variable with eigenstates $\vert x_\theta \rangle$ corresponding
to the eigenvalues $x_\theta \in \mathbb R$, i.e.,
$\hat X_\theta \vert x_\theta \rangle  = x_\theta  \vert x_\theta \rangle$.

Homodyne detection is a diffusive type of measurement
dominated by Gaussian noise.
For a fixed value of $\theta$, through repeated measurements
we can estimate the quadrature probabilities
${\rm pr}_\rho(x_\theta) = \langle x_\theta \vert \rho \vert x_\theta \rangle$.
From this data,
the state $\rho$ can be inferred via Maximum Likelihood estimation~\cite{Lvovsky2009}.

Since $\hat X_\theta$ is a continuous variable and the total
number of measurements
is finite, the data must be binned.
In order to do this, we split
a relevant domain of the real axis into intervals $[x_j, x_{j+1}]$.
From here, the binned quadrature probabilities, for fixed $\theta$, are obtained
by integration as
\begin{equation}
{\rm pr}_{\rho}(\theta, j)  = \int_{x_j}^{x_{j + 1}} {\rm d}x \, \langle x_\theta  \vert  \rho \vert x_\theta  \rangle .
\end{equation}
Maximum Likelihood is a statistical inference method.
The idea is to search for the physical state which maximizes the probability
of obtaining the measured quadrature probabilities.

For an arbitrary state $\widetilde \rho$, we define the \emph{likelihood function}
\begin{equation}
L(\widetilde \rho) = \prod_{\theta, j} \left[ {\rm pr}_{\widetilde \rho}(\theta, j) \right]^{n_{\theta, j} },
\end{equation}
with $n_{\theta, j}$ the number of measurements in the $j$-th bin, i.e.,
the binned histogram. The latter is related to the actual state $\rho$ 
we are trying to infer.
The goal is to find the state $\rho^*$ which maximizes the likelihood function.
This state will be our best approximation to  $\rho$.

The state $\rho^*$ which maximizes the likelihood obeys the extremal equation
 \begin{equation}\label{app:extr}
\hat R(\rho^*)\rho^* \hat R(\rho^*)=\rho^*,
 \end{equation}
 with the state-dependent operator $\hat R$ defined as
  \begin{equation}
  \hat R(\widetilde \rho) = \frac{1}{n} \sum_ {\theta} n_\theta \sum_j \frac{(n_{\theta,j}/ n_\theta)}{\text{pr}_{\widetilde  \rho}(\theta, j)} \hat \Pi_{\theta,j }.
 \end{equation}
Here $n_\theta = \sum_j n_{\theta, j}$ is the total number of measurements for
a fixed phase $\theta$,
$n = \sum_\theta n_\theta$ is the total number of measurements and $\hat \Pi_{\theta,j }$ the projector
into the $j$-th bin
\begin{equation}\label{app:proj}
\hat \Pi_{\theta,j } = \int_{x_j}^{x_{j + 1}} {\rm d}x \, \vert x_\theta \rangle \langle x_\theta \vert .
\end{equation}

The extremal  condition \eqref{app:extr} amounts to $\hat R(\rho^*)$ being proportional
to the identity operator. Indeed, for $\rho^* = \rho$ the probability
${\rm pr}_{\rho^*}(\theta, j)$ 
equals 
the $j$-th entry of the
normalized measurement
histogram $n_{\theta,j}/ n_\theta$, and therefore, we are left
with a sum of projectors over the different bins. Following Eq. \eqref{app:proj},
the sum is equal to the identity operator.

The state which maximizes the likelihood function can be found iteratively.
First, we choose a basis in which to represent the density matrix. In our case,
this will be the Fock or number basis with a photon number cutoff $N_{\rm Fock}$.
Then,
starting from an arbitrary state, in our case $\rho_0 =  \mathbb 1 /  N_{\rm Fock}$,
the iterative search proceeds as follows
\begin{equation}\label{iteq}
  \rho_{k+1} = \mathcal N [R(\rho_k)\rho_k R(\rho_k)],
\end{equation}
where $\mathcal N$ denotes normalization to unity trace.
We stop the iterations when the variation in the state,
quantified by a suitable norm,
 for consecutive
iterations
is sufficiently small. In our case,
\begin{equation}\label{deltarho}
\Delta \rho = \lVert \rho_{k+1}-\rho_k \rVert \le 10^{-6}.
\end{equation}
The chosen matrix norm is the Frobenius norm, defined as $\lVert
A\lVert = \sqrt{\text{Tr}[A^\dagger A]}$ for a matrix $A$.

\end{widetext}

\end{appendix}

\end{document}